\documentclass[aps,pra,letterpaper,twocolumn,amsmath,amssymb,nofootinbib,superscriptaddress
,floatfix
]{revtex4}

\usepackage{mathtools}
\usepackage{booktabs}
\usepackage{slashed}
\usepackage{bm}

\newcommand{\bra}[1]{\langle#1\rvert}
\newcommand{\ket}[1]{\lvert#1\rangle}
\newcommand{\overlap}[2]{\langle{#1}|{#2}\rangle}

\newcommand{\ketbra}[2]{|{#1}\rangle\langle{#2}|}
\newcommand{\braket}[2]{\langle{#1}|{#2}\rangle}

\newcommand{\smallfrac}[2]{\mbox{$\frac{#1}{#2}$}}

\newcommand{\gdisp}{\slashed{g}}
\newcommand{\op}[1]{\hat{#1}}
\renewcommand{\bra}[1]{\left\langle#1\right\rvert}
\renewcommand{\ket}[1]{\left\lvert#1\right\rangle}
\renewcommand{\ketbra}[2]{\left\lvert{#1}\middle\rangle\middle\langle{#2}\right\rvert}
\renewcommand{\braket}[2]{\left\langle{#1}\middle\vert{#2}\right\rangle}
\newcommand{\inprod}{\braket}

\newcommand{\choices}[2]{\genfrac{\lbrace}{\rbrace}{0pt}{}{#1}{#2}}
\def\negspace{\!}

\def\rsub#1#2{{#1} \negspace {\protect\vphantom{#1}}_{#2}}
\def\lrsub#1#2#3{{\protect\vphantom{#1}}_{#2} \negspace {#1} \negspace {\protect\vphantom{#1}}_{#3}}

\def\ketsub#1#2{\rsub {\ket{#1}} {#2}}

\def\pket#1{\ketsub{#1} p}
\def\qket#1{\ketsub{#1} q}

\def\inprodsubsub#1#2#3#4{\lrsub {\inprod{#1}{#2}} {#3} {#4}}

\newcommand{\abs}[1]{\left\lvert{#1}\right\rvert}

\renewcommand{\Re}{\operatorname{Re}}

\newcommand{\approptoinn}[2]{\mathrel{\vcenter{
  \offinterlineskip\halign{\hfil$##$\cr
    #1\propto\cr\noalign{\kern2pt}#1\sim\cr\noalign{\kern-2pt}}}}}

\newcommand{\appropto}{\mathpalette\approptoinn\relax}
\newcommand{\simapprox}{\appropto}

\def\integers{\mathbb{Z}}

\def\id{I}

\def\1{\mat{\id}}

\def\mat#1{\bm{\mathrm{#1}}}

\newcommand{\logic}{\text{L}}

\newcommand{\dsign}{\varsigma}

\newcommand{\logicid}{\text{L,ideal}}
\newcommand{\sqp}{r}
\newcommand{\psiresource}{ \psi_{\pm}^{\rm res} }

\newcommand{\NL}{{N_L}}
\newcommand{\NA}{{N}}
\newcommand{\emberr}{\xi}

\usepackage{graphicx}
\usepackage{mathrsfs}
\usepackage[usenames,dvipsnames]{xcolor}
\usepackage{hyperref}
\hypersetup{colorlinks = true, linkcolor = [rgb]{0.19411,0.51882,0.667058}, urlcolor = [rgb]{0.125490,0.29542,0.1647058}, citecolor = [rgb]{0.75882,0.37411,0.14117}}

\usepackage{soul}

\graphicspath{{graphics/}}

\allowdisplaybreaks

\begin{document}

\bibliographystyle{apsrev}

\title{Encoding qubits into oscillators with atomic ensembles and squeezed light}

\author{Keith R. Motes}
\email[]{motesk@gmail.com}
\affiliation{Centre for Engineered Quantum Systems, Department of Physics and Astronomy, Macquarie University, Sydney, New South Wales 2113, Australia}

\author{Ben Q. Baragiola}
\email[]{ben.baragiola@gmail.com}
\affiliation{Centre for Engineered Quantum Systems, Department of Physics and Astronomy, Macquarie University, Sydney, New South Wales 2113, Australia}

\author{Alexei Gilchrist}
\email[]{alexei@entropy.energy}
\affiliation{Centre for Engineered Quantum Systems, Department of Physics and Astronomy, Macquarie University, Sydney, New South Wales 2113, Australia}

\author{Nicolas C. Menicucci}
\email[]{ncmenicucci@gmail.com}
\affiliation{Centre for Quantum Computation and Communication Technology, School of Science, RMIT University, Melbourne, Victoria 3001, Australia}
\affiliation{School of Physics, The University of Sydney, Sydney, New South Wales 2006, Australia}

\date{\today}

\begin{abstract}
The Gottesman-Kitaev-Preskill (GKP) encoding of a qubit within an oscillator provides a number of advantages when used in a fault-tolerant architecture for quantum computing, most notably that Gaussian operations suffice to implement all single- and two-qubit Clifford gates. The main drawback of the encoding is that the logical states themselves are challenging to produce. Here we present a method for generating optical GKP-encoded qubits by coupling an atomic ensemble to a squeezed state of light. Particular outcomes of a subsequent spin measurement of the ensemble herald successful generation of the resource state in the optical mode. We analyze the method in terms of the resources required (total spin and amount of squeezing) and the probability of success. We propose a physical implementation using a Faraday-based quantum non-demolition interaction.
\end{abstract}

\maketitle

\section{Introduction}
\label{sec:intro}

Noise is ubiquitous in physical devices, and quantum devices---including quantum computers~\cite{bib:NielsenChuang00}---are no exception. Fortunately, the threshold theorem states that, under some reasonable assumptions about the type of noise, as long as the level of that noise is below a nonzero minimum value called the \emph{fault-tolerance threshold}, quantum error correction can be used to reduce them to arbitrarily low levels (see Ref.~\cite{Gottesman:2009ug} for a review). Quantum error correction relies on three key components: (1)~redundantly encoded quantum information, (2)~a method to detect and correct errors, and (3)~a way to prepare, manipulate, and measure the quantum information in its encoded form.

In 2001, Gottesman, Kitaev, and Preskill~(GKP) proposed a method to encode a qubit into a harmonic oscillator in a way that protects against small displacement errors~\cite{Gottesman2001}. In fact, this protection is fully general since any error acting on the oscillator can be expanded as a superposition of displacements~\cite{Gottesman2001}. The main advantage of this scheme is that all logical Clifford operations---the most important and most common qubit gates~\cite{bib:NielsenChuang00}---are implemented through Gaussian operations at the physical level. This is an advantage for optical implementations because Gaussian operations are much easier to implement than non-Gaussian ones~\cite{Braunstein2005a}. On the other hand, the logical states are ideally an infinite comb of $\delta$~functions (i.e.,~a Dirac comb), which are impossible to produce because such states are unphysical (requiring infinite energy). As such, approximate states must be used instead.

A comb of narrow Gaussian spikes whose heights are distributed according to a wide Gaussian envelope serves as the usual approximate wave function for the logical qubits, carrying with it an associated ``embedded error''~\cite{Gottesman2001} because it does not lie fully within the ideal logical subspace. As the spikes get ever narrower and the envelope of heights ever wider, these approximate states tend to their ideal counterparts. Even in their approximate form, GKP states are notoriously difficult to make in the laboratory, despite numerous proposals to do so~%
\cite{Gottesman2001,Vasconcelos:2010gb,bib:Travaglione02,Pirandola:2004jo,Pirandola:2006gh,Pirandola:2006bf,Terhal2016}.

Recent years have seen renewed interest in conquering these difficulties and producing GKP-encoded qubits precisely because of the ability to implement qubit-level Clifford gates as Gaussian operations. Specifically, they dovetail well with continuous-variable~(CV) measurement-based quantum computing, in which all Gaussian operations can be implemented simply with homodyne detection on a pre-made CV cluster state~\cite{Menicucci2006}. This architecture is particularly promising because optical CV cluster states can be made on an exceptionally large scale~\cite{Yokoyama:2013jp,Chen:2014jx, Yoshikawa2016} with minimal experimental equipment~\cite{Menicucci2011a,Wang:2014im,Alexander2016}. The price one pays for this unprecedented scalability is ubiquitous noise due to finite squeezing~\cite{Alexander:2014ew}. Nevertheless, this noise can be handled---and fault-tolerant quantum computing achieved~\cite{Menicucci:2014cx}---by processing GKP-encoded qubits~\cite{Gottesman2001} and applying standard quantum error correction techniques~\cite{Gottesman:2009ug}.

In this work, we propose a method to generate optical GKP states by coupling an atomic ensemble, prepared in a spin coherent state, to a squeezed state of light. In Sec.~\ref{sec:GKPanatomy}, we briefly review GKP states and their error-correction properties, and we identify the target states of our proposal. In Secs.~\ref{sec:lightensemble} and~\ref{sec:stategen}, we describe the light-matter interaction and show that the resulting optical states match the target states. In Sec.~\ref{sec:requirements}, we analyze the method in terms of its parameters and success probability as a function of output-state quality. In Sec.~\ref{sec:PhysicalImplementation}, we discuss the experimental requirements to implement the method using a Faraday interaction with an atomic ensemble. Section~\ref{sec:conclusion} concludes.

\section{Ideal and approximate GKP states}
\label{sec:GKPanatomy}
\subsection{Ideal GKP states and correcting errors}

The logical subspace of the GKP-encoded qubit is spanned by the logical states~$\{\ket 0_\logicid, \ket 1_\logicid\}$, where
\begin{align}
\label{eq:idealGKP}
	\ket j_\logicid \propto \sum_{s = -\infty}^\infty \qket {(2s+j) g} \propto \sum_{s = -\infty}^\infty (-1)^{js} \pket {s\pi/g},
\end{align}
where $\qket u$ and $\pket u$ are, respectively, position and momentum eigenstates (identified by the subscript on the ket), which satisfy $\op q \qket u = u \qket u$ and $\op p \pket u = u \pket u$. The wave function for $\ket j_\logicid$ is a Dirac comb in position with spacing~$2g$ and offset $jg$ or, equivalently, a superposition of two Dirac combs in momentum, each with spacing~$2\pi/g$, offset from each other by~$\pi/g$, and with a relative phase of $(-1)^j$. Note that we have omitted the normalization because these ideal logical states are not normalizable but can nevertheless be approximated arbitrarily well by physical states. (We discuss approximate states below.)

GKP states are protected against shift errors in position and/or momentum~\cite{Gottesman2001}. The peaks in the position-space wave function of the logical $\ket{0}_\logicid$ and $\ket{1}_\logicid$ states are interleaved such that the closest distance between peaks of one state and those of the other is~$g$. If any logical GKP state $\ket \psi = \alpha \ket 0_\logicid + \beta \ket 1_\logicid$ is shifted by a small amount $\delta q$ in position, then this shift error can be corrected by measuring $\op q \mod g$ (by coupling to an ancilla prepared in~$\ket 0_\logicid$) and then shifting the state back into the code space~\cite{Gottesman2001}. When $\abs{\delta q} < g/2$, this correction succeeds perfectly. But when $g/2 < \abs{\delta q} < g$, the attempt at correction results in a Pauli-$X$ error on the logical state (since it is more likely that the given measurement of $\op q \mod g$ would have resulted from a shift by $\delta q - g$ rather than by $\delta q$). Similar properties hold for momentum shifts~$\delta p$, which are correctable when $\abs{\delta p} < \pi/2g$ and result in a Pauli-$Z$ error for larger shifts~\cite{Gottesman2001}. 

Since displacements in position and momentum (Weyl operators) form a complete basis for expanding any operator, the ability to correct shift errors is tantamount to protection against \emph{any} (small) error~\cite{Gottesman2001}, including common ones like amplitude damping (photon loss). The syndrome measurement procedure prescribed in Ref.~\cite{Gottesman2001} projects the more general error into definite shifts in position and momentum, which are then (approximately) corrected. This ability to project more general errors into a basis of correctable ones through syndrome measurements is the critical feature of any scheme of active quantum error correction~\cite{Gottesman:2009ug}.

\subsection{Approximate GKP states}

Both in the original work of GKP~\cite{Gottesman2001} and here, when we talk of a ``GKP state'' we mean a physical state approximating a state in the GKP logical subspace. Given an ideal GKP state~$\ket \psi$, we will limit ourselves to considering approximate states of the form
\begin{align}
\label{eq:psieta}
	\ket \psi_\emberr \propto \op \emberr \ket \psi = \int du\, dv\, \emberr(u, v) e^{-iu\op p + iv\op q} \ket \psi,
\end{align}
where $\op \emberr$ is a member of the \emph{affine oscillator semigroup}~\cite{Howe1988} (we will come back to this), and $\emberr(u,v)$ is the Weyl-operator expansion of~$\op \emberr$. We assume that $\op \emberr$ is close to the identity so that $\ket \psi_\emberr$ is a good approximation to~$\ket \psi$.

Without ambiguity, we will refer to both $\op \emberr$ and $\emberr(u,v)$ as the \emph{embedded error} in the approximate GKP state. This terminology is consistent with Ref.~\cite{Gottesman2001} and makes sense because we can think of $\ket \psi_\emberr$ as an ideal GKP state~$\ket \psi$ followed immediately by a trace-decreasing CP map (an error), described by the single Kraus operator~$\op{\xi}$.

As a GKP state is used for computation, its embedded error will get larger, but this can be partially corrected if one has access to ancillary systems prepared in a (still approximate) logical basis state~\cite{Gottesman2001}. It is easiest to see this by noting that $\op \emberr$, like any operator, can be written as a superposition of shifts in position and momentum, Eq.~\eqref{eq:psieta}, which gets projected to just a single shift during the syndrome measurement and then approximately corrected, as discussed above and in Ref.~\cite{Gottesman2001}. This is why it is important to have a supply of high-quality approximations to GKP basis states: a fresh ancilla is needed each time one wants to do GKP error correction.

We consider slightly more general embedded errors than GKP themselves did because of the semigroup we have allowed $\op \emberr$ to inhabit. The mathematical details of this semigroup (which can be found in~\S18 of~\cite{Howe1988}) are rather complicated and beyond the scope of this work, so instead we will merely highlight important features and the way our states go beyond those considered by GKP.

In their original work~\cite{Gottesman2001}, the most general embedded error considered by GKP is
\begin{align}
\label{eq:etaGKP}
	\emberr(u,v) 
	\propto
	\exp \left( -\frac {u^2} {2\Delta^2} - \frac {v^2} {2\kappa^2}  \right),
\end{align}
which, for $\Delta \ll 1$ and $\kappa \ll 1$ (and appropriate normalization), corresponds to
\begin{align}
\label{eq:etaopGKP}
	\op \emberr
	&
	\approx e^{-\Delta^2 \op p^2/2} e^{-\kappa^2 \op q^2/2}
	\approx e^{-\kappa^2 \op q^2/2} e^{-\Delta^2 \op p^2/2}
	.
\end{align}
An operator of the form $e^{-\sigma^2 \op q^2/2}$ acts on a position-space wave function~$\psi(q)$ as multiplication by a Gaussian envelope with zero mean and variance~$1/\sigma^2$ [i.e., $\psi(q) \mapsto e^{-\sigma^2 q^2/2} \psi(q)$]. The equivalent action of this operator on a momentum-space wave function~$\tilde \psi(p)$ is to %
convolve it with a zero-mean Gaussian with variance~$\sigma^2$ [i.e., $\tilde \psi(p) \mapsto (2 \pi  \sigma ^2)^{-1/2} \int d\tau\, {\tilde \psi(p - \tau)} e^{-\tau^2/2\sigma^2}$], resulting in a blurred version of the original wave function. The operator~$e^{-\sigma^2 \op p^2/2}$ behaves exactly the same, except with the roles of position and momentum exchanged.%

When the parameter $\sigma \ll 1$, these two operators approximately commute, and therefore we can regard $\op \emberr$ in Eq.~\eqref{eq:etaopGKP} as applying a large Gaussian envelope in position with variance~$1/\kappa^2$ and replacing each spike in position with a narrow Gaussian with variance~$\Delta^2$, without caring too much about the order in which these happen. This is the type of approximate state originally considered by GKP~\cite{Gottesman2001}.

Operators of the form~$e^{-\sigma^2 \op q^2/2}$ or~$e^{-\sigma^2 \op p^2/2}$ are members of the (ordinary) oscillator semigroup~\cite{Howe1988}. This semigroup has a nontrivial multiplication law, which is why we have restricted discussion to the simpler special case~$\sigma \ll 1$. This restriction nevertheless allows us to represent all possible combinations of large, zero-mean, Gaussian envelopes applied in position and/or momentum.

So why is this not enough? In other words, why must we consider any other states beyond just these? There are two reasons. The first is that the main application we have in mind for these states---fault-tolerant CV measurement-based quantum computing~\cite{Menicucci:2014cx}---has ubiquitous displacements that must be corrected~\cite{Menicucci2006}, which, combined with intrinsic error at each step due to finite squeezing~\cite{Alexander:2014ew,Menicucci2006}, means that even if we start with this type of state, we will very quickly end up with one outside of this class. The second reason is that the generation procedure we describe below sometimes produces states that are outside of this class
 (see Sec.~\ref{sec:stategen}).

Specifically, we extend the embedded errors to include those that arise when $\op \emberr$ is a member of the \emph{affine} oscillator semigroup~\cite{Howe1988}---which includes the ordinary one as a sub-semigroup. This allows us to represent all possible combinations of large, \emph{arbitrary-mean}, Gaussian envelopes applied in position and/or momentum. We want this because the ideal GKP codespace is periodic in position and in momentum, so there is no fundamental reason to prefer zero-mean envelopes to more general ones, and these more general states arise naturally anyway, as discussed above. Importantly, note that applying these envelopes does not shift the locations of the spikes when applied to an ideal GKP state; they merely enlarge the class of embedded errors that can be discussed.

\subsection{Logical states, target states, resource states, and specifying their quality}
\label{sec:GKPtarget}

Having discussed what it means for a state to be an approximate GKP state and how we have expanded this definition beyond that discussed originally by GKP~\cite{Gottesman2001}, we now identify the particular parameters of the target approximate GKP states we wish to produce.

The most important motivating factor in this decision is the main application we have in mind: fault-tolerant CV measurement-based quantum computing~\cite{Menicucci:2014cx}. In this scenario, the actual errors that accumulate in the GKP states will be roughly symmetric in position and momentum over the course of a computation. Therefore, we want an encoding that is unbiased with respect to position and momentum.

This involves two requirements. First, we require
\begin{align}
	g = \sqrt \pi
\end{align}
so that the spacing between spikes in the grid comprising the logical subspace is equal for both position and momentum---see Eq.~\eqref{eq:idealGKP}, and choose $g = \pi/g$. Second, we want a logical state whose embedded error is roughly symmetric in position and momentum:
\begin{align} \label{eq:emberreq}
	\sigma
	\coloneqq
	\Delta = \kappa
	.
\end{align}

For the rest of this article we focus on approximate GKP states $\ket{j}_\logic  = \op \emberr \ket{j}_\logicid$, which have embedded errors of the form
\begin{align}
\label{eq:emberrspec}
	\op \emberr
	&
	\approx e^{-\sigma^2 \op p^2/2} e^{-\sigma^2 (\op q-q_0)^2/2}
\nonumber
\\
	&
	\approx e^{-\sigma^2 (\op q-q_0)^2/2} e^{-\sigma^2 \op p^2/2}
	,
\end{align}
for small $\sigma$. The momentum-space envelope always has a mean of zero, but that of the position-space envelope may be nonzero (see Sec.~\ref{sec:stategen}).
 The possibility of a nonzero mean in the position-space envelope is the key difference between the target states we study here and those analyzed in depth by GKP. This difference is largely inconsequential, however, because the resulting states are all decent approximations to ideal states when the applied envelopes are large.
 
Henceforth, we refer to $\ket{j}_\logic$ (and superpositions thereof) as \emph{logical states}, keeping in mind that they include the embedded error as described above.
For reasons that will become clear in Sec.~\ref{sec:stategen}, we focus on creating Pauli-$X$-basis logical states~$\ket \pm_\logic = {\frac {1} {\sqrt 2} \bigl(\ket 0_\logic \pm \ket 1_\logic \bigr)}$, which we refer to as \emph{target states}. 

\begin{figure*}[t]
\includegraphics[width=.9\textwidth]{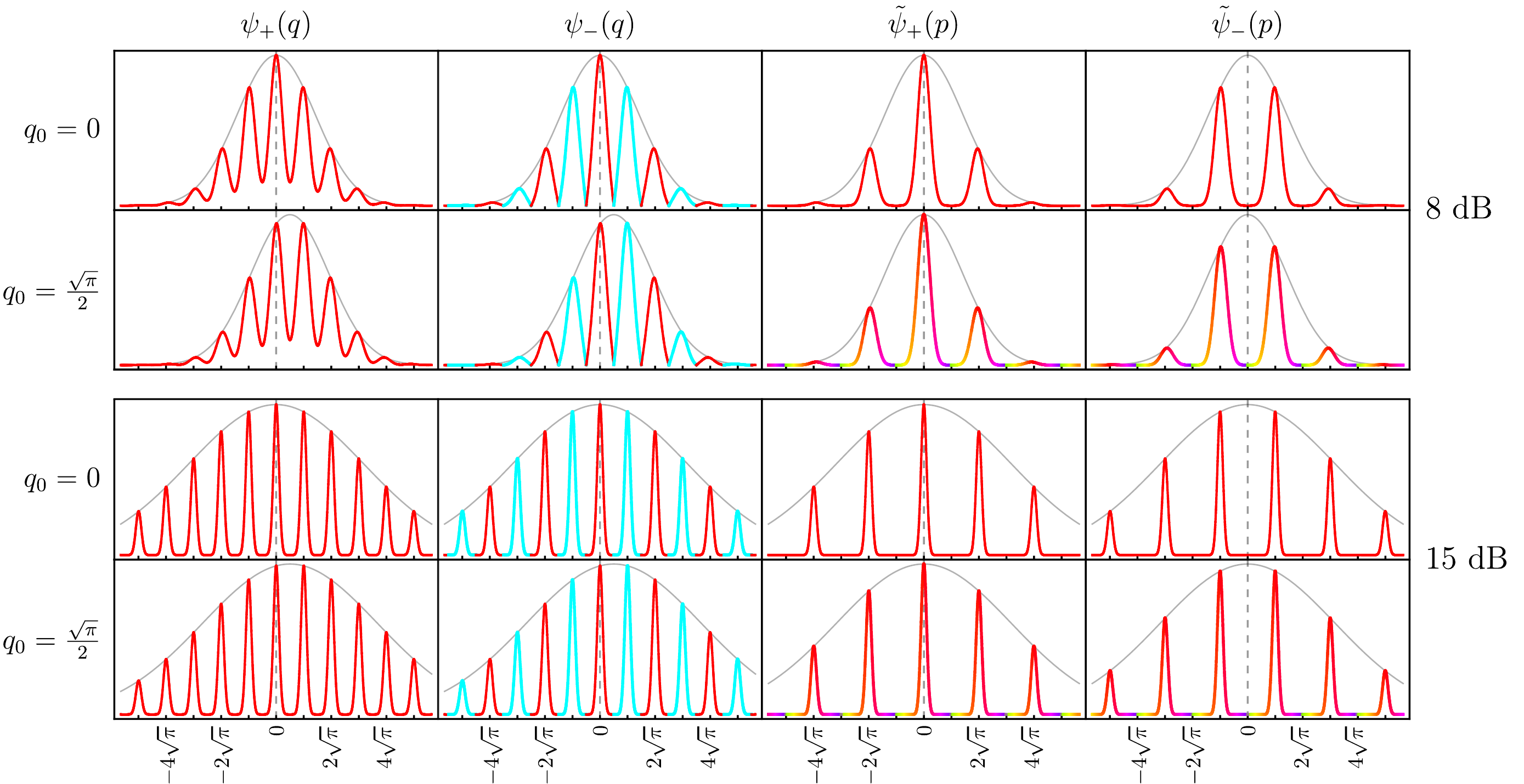}%
\hfill%
\llap{\raisebox{9.4em}{\includegraphics[width=.13\textwidth]{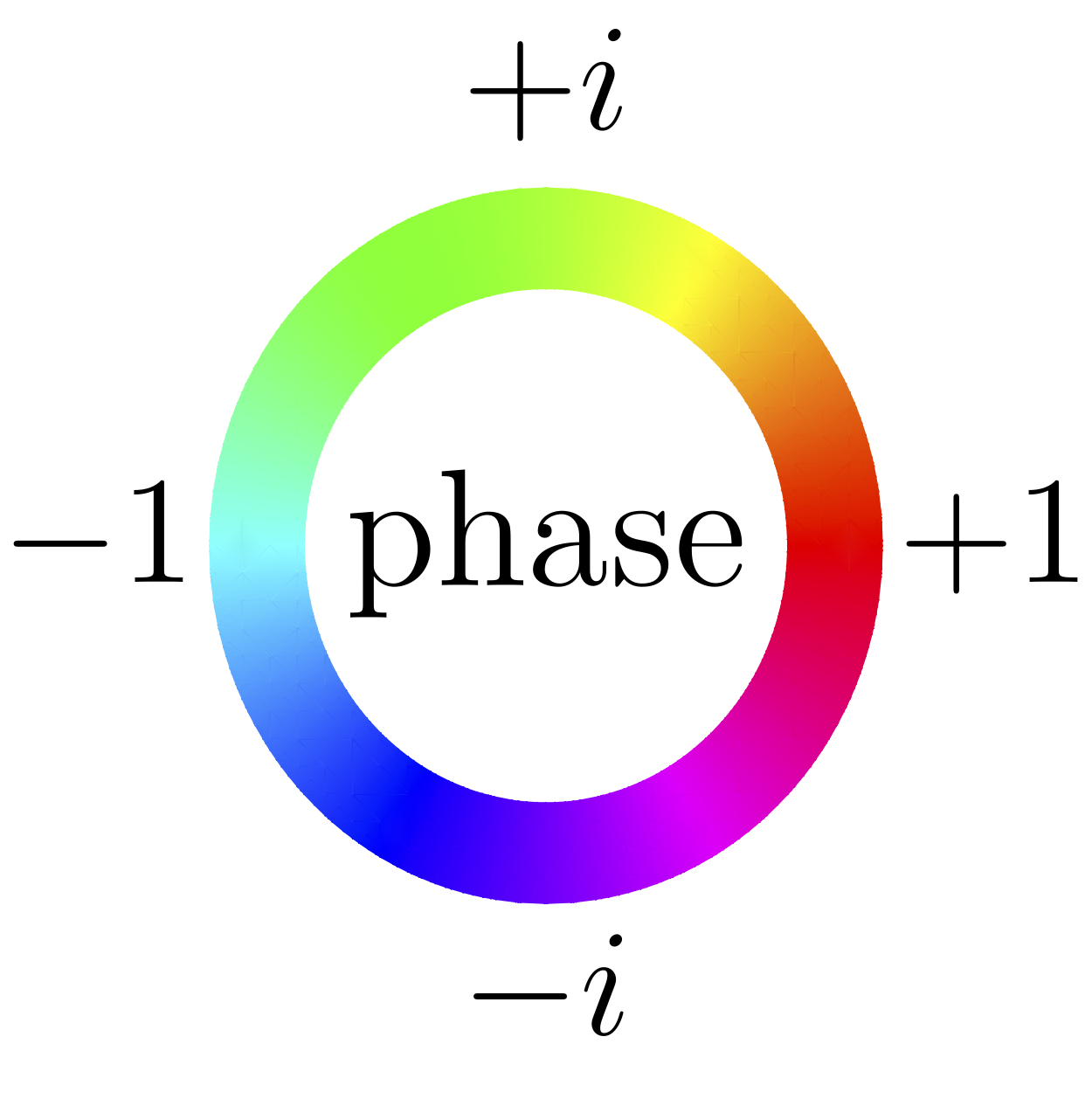}}}\ 
\caption{Example wave functions [Eqs.~\eqref{eq:targetq} and~\eqref{eq:targetp}] for low-quality (8-dB) and high-quality (15-dB) target states. These states all have equal spacing in momentum and position ($g=\sqrt \pi$) and symmetric embedded error, allowing us to express their quality using a single value---the squeezing of the state, as defined in Eq.~\eqref{eq:squeezingofstate}. This parameter determines the width of the spikes and envelopes through Eq.~\eqref{eq:sigmafromsqueezing}. %
The red and cyan (darker and lighter) in~$\psi_-(q)$ indicate a relative phase of $\pm 1$, respectively, alternating between adjacent spikes, while the phase is uniform for~$\psi_+(q)$. Notice that when the position-space envelope (thin, gray) is not centered (i.e.,~$q_0 \neq 0$), the momentum-space representation has a nonconstant phase, indicated by color (greyscale variation) as shown in the legend on the right. This variation is nontrivial only where the magnitude of the wavefunction is negligible and is therefore largely immaterial to the error-correction properties of the state because the spikes themselves---i.e., the high-amplitude parts of the state---all have the appropriate location and phase in both position and momentum.%
\label{fig:SymmResourceState}
}
\end{figure*}

We will present the target states in terms of their wave functions. 
We denote the position- and momentum-space wave functions of a state~$\ket\psi$, respectively, by
\begin{align}
	\psi(u) \coloneqq \inprodsubsub u \psi q {}
	\qquad \text{and} \qquad
	\tilde\psi(u) \coloneqq \inprodsubsub u \psi p {}
	,
\end{align}
where the notation for position and momentum eigenstates is defined below Eq.~\eqref{eq:idealGKP}. These are related by the Fourier transform
\begin{align}
\label{eq:}
	\tilde \psi (p)
	&
	=
	\frac {1} {\sqrt {2\pi}}
	\int dq\, e^{-i q p} \psi (q).
\end{align}
Our target states have the position-space wave function
\begin{align}
\label{eq:targetq}
	\psi_{\pm} (q) 
	&
	\simapprox
	e^{-\sigma^2 (q-q_0)^2/2} \sum_{s \in \integers} (\pm 1)^s \exp\left[-\frac {(q-s\sqrt \pi)^2} {2\sigma^2} \right]
	,
\end{align}
with $\simapprox$ indicating both approximation (valid for $\sigma \ll 1$) and proportionality up to a constant. We also require
$\abs{q_0} \le \sqrt \pi/2$ because if $\abs{q_0}$ were larger than this, we could reduce it to the indicated range by shifting the state in position by an integer multiple of~$\sqrt \pi$.
The equivalent momentum-space wave functions are (approximately) \begin{align}
\label{eq:targetp}
	\tilde \psi_{\pm} (p) 
	&\simapprox
	e^{-\sigma^2 p^2/2} \sum_{\mathclap{s \in \choices {2\integers\phantom{+1}} {2\integers+1}}} e^{-i q_0 (p-s\sqrt \pi)} \exp\left[- \frac{(p-s\sqrt \pi)^2}{2 \sigma ^2} \right]
	,
\end{align}
where the upper and lower sets (evens and odds) correspond to $\pm$, respectively. Several examples of these wave functions are shown in Fig.~\ref{fig:SymmResourceState}. 

Instead of directly calculating the Fourier transform of Eq.~\eqref{eq:targetq}---which, with $\sigma \ll 1$, would ever so slightly shift the spike locations---we have elected to use the second form of~$\op \emberr$ from Eq.~\eqref{eq:emberrspec} and apply it directly to the ideal momentum-space representation of~$\ket \pm_\logicid$. This results in a simpler form of the approximate wave function and is valid for our purposes because our target state is one with $\sigma \ll 1$.

Our goal is to produce high-quality %
physical approximations to ideal GKP states
for use in CV measurement-based quantum computing. To be useful for this purpose, such a state---assumed pure for the present discussion---must have a wave function that closely approximates one of the target states~$\ket{\pm}_\logic$ described above%
. We henceforth refer to any such state as a \emph{resource state}.

Since the logical states, target states, and resource states under consideration all have symmetric embedded error [Eq.~\eqref{eq:emberrspec}], we can measure the quality of any of these states by the single parameter, $\sigma^2$, which represents the variance of an individual spike in the superposition. We can express this parameter in~dB, and when we do so, we call this the \emph{squeezing} of the state. It is a measure of overall quality of the state because it is the only parameter required to specify the embedded error (up to shifts in the envelope). And it has operational value because of its connection to the amount of optical squeezing that would be required to produce just one of the spikes in the superposition (see Sec.~\ref{sec:lightensemble}). Specifically,
\begin{align}
\label{eq:squeezingofstate}
	\text{(squeezing of the state in dB)} \coloneqq -10 \log_{10} \sigma^2.
\end{align}
Inverting this relation specifies the embedded error [through Eq.~\eqref{eq:emberrspec}] and has the following visual interpretation with respect to the state's wave function, as shown in Fig.~\ref{fig:SymmResourceState}:
\begin{align}
\label{eq:sigmafromsqueezing}
	\text{(envelope width)}^{-1} = \text{(spike width)} = \sigma = 10^{-\text{(\#dB)}/20},
\end{align}
where (\#dB) is the squeezing of the state in dB.

\section{Light-ensemble interaction}
\label{sec:lightensemble}

\begin{figure}[tb]
\includegraphics[width=.9\columnwidth]{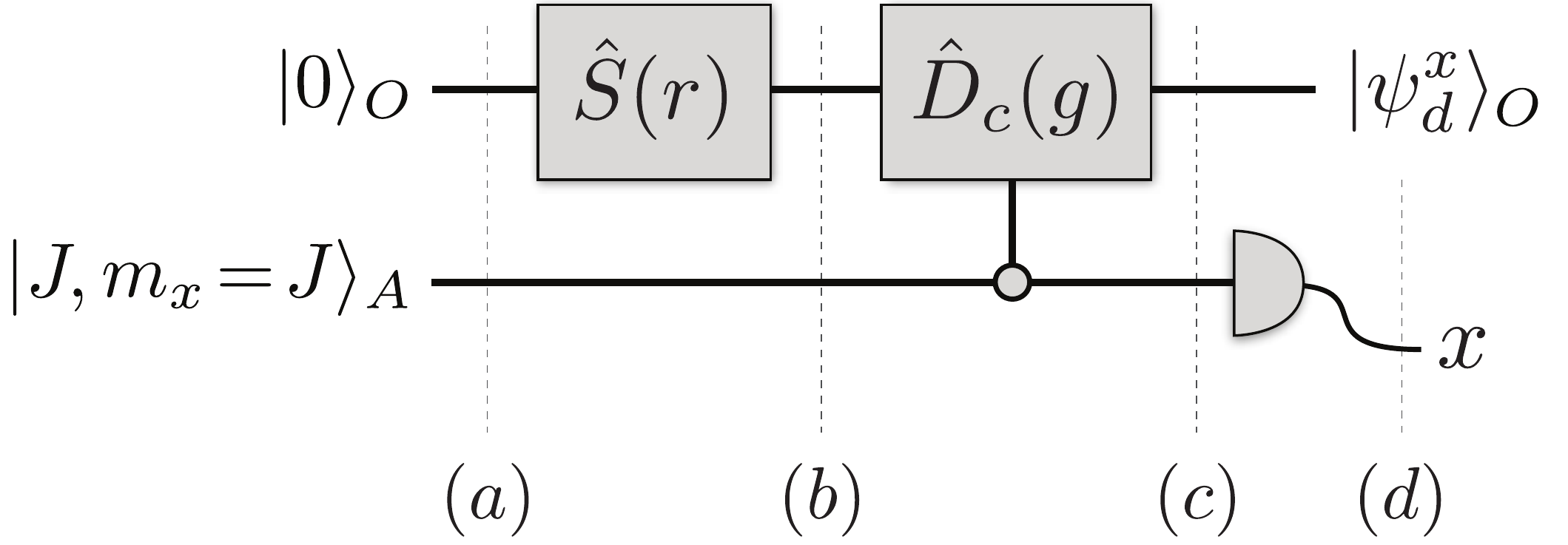}
\caption{ Circuit that creates a resource state. The optical mode is prepared in the vacuum state~$\ket{0}_O$, and an atomic ensemble, with total collective spin $J$, is prepared in a spin-coherent state polarized along $x$, $\ket{J, m_x=J}_A$. The optical mode is first squeezed via the standard squeezing operator~$\op{S}(\sqp)$ and is then coupled to the collective spin via a controlled position shift~$\op{D}_c(g)$. The spin is then projectively measured yielding outcome $x$, conditionally preparing the optical mode in state $\ket{\psi_d^x}_O$. When certain conditions are met (see Sec. \ref{sec:stategen}), this state may serve as a resource state, as defined at the end of Sec.~\ref{sec:GKPtarget}.\label{fig:CircuitModel}}
\end{figure}

In this section we describe how resource states can be conditionally prepared by entangling an atomic ensemble and an optical mode followed by projective measurement of the collective atomic spin. Where appropriate, the spin state of the atomic ensemble and the optical state of the field mode are labelled with $A$ and $O$, respectively. This section follows the progress of the joint spin-optical state through the circuit in Fig.~\ref{fig:CircuitModel}, which begins with an unentangled state $\ket{\Psi_a}_{AO}=\ket{J,m_x\!=\!J}_A \otimes \ket{0}_O$ at (a) and ends with a conditional optical state prepared upon a measurement of the spin at (d). A succinct derivation of the protocol for general initial angular-momentum and optical states is given in Appendix~\ref{sec:GKPKraus}.  

Initially the atomic ensemble with total collective spin~$J$ is prepared in a spin-coherent state %
polarized along the $x$-axis, which we will denote as $\ket{J, m_x\!=\!J}_A $. 
Expressed in terms of eigenstates along the $z$-axis, $\ket{J,m_z}_A$, the initial spin state is
\begin{align}
  \ket{J,m_x\!=\!J}_A &= \op{R}_y(\pi/2) \ket{J,m_z\!=\!J}_A \nonumber \\ 
  &= \sum_{\mathclap{m=-J}}^J d_{m,J}^{(J)}(\pi/2) \ket{J,m}_A,
\end{align}
where the coefficients ${d_{m,m'}^{(J)}(\beta) \coloneqq \bra{J,m} \op{R}_y(\beta) \ket{J,m'}}$, known as Wigner's (small) $d$-matrix elements~\cite{Varshalovich1988}, are  matrix elements of the operator $\op{R}_y(\beta)=\exp(-i \beta \op{J}_y)$, which rotates the spin state around the $y$-axis by an angle~$\beta$. For our purposes, $J$ is fixed and $\beta=\pi/2$; henceforth, we drop unnecessary notation and define
\begin{align}
\label{eq:dmmd}
d_{m,m'} &\coloneqq d_{m,m'}^{(J)}\left(\frac \pi 2 \right) %
= \frac{1}{2^J} \sum_{k} (-1)^{k-m'+m} \nonumber \\
&\quad \times \frac{\sqrt{(J+m')!(J-m')!(J+m)!(J-m)!}}{(J+m'-k)!k!(J-k-m)!(k-m'+m)!},
\end{align}
where the sum is only over integers~$k$ such that no factorial in the denominator is of a negative number. We have used the explicit form for $d_{m,m'}^{(J)}(\beta)$ that appears, $e.g.$, in Sakurai~\cite{bib:sakurai2011modern} Sec.~3.8,
but this expression can in fact be summed analytically and written in terms of Jacobi polynomials~\cite{Varshalovich1988}. We include these details in Appendix~\ref{app:analyticform} and use it to derive two special cases used in the main text.

One of these special cases, however, is simple enough to derive directly from the explicit sum, Eq.~(\ref{eq:dmmd}):
When $m'=\pm J$, only a single term in the sum has non-negative factorials in the denominator, and the matrix element simplifies to
\begin{align}
\label{eq:GKPmatrixElements}
  d_{m,\pm J} = \frac{(\pm 1)^{J+m}}{2^J}\begin{pmatrix} 2J \\ J-m\end{pmatrix}^{1/2}.
\end{align}
(See Appendix~\ref{app:analyticform} for an alternative derivation.) Specifically, the factorials in the denominator are non-negative only when $k=J-m$ for $m'=+J$; and when $k=0$ for $m'=-J$. The values of $m'=\pm J$ are significant as these yield desirable resource states, as we will see in Sec.~\ref{sec:stategen}.

The initial optical state at~(a) is the bosonic vacuum~$\ket{0}$ with creation and annihilation operators $\op{a}$ and $\op{a}^\dagger$ obeying the usual commutation relations $[\op{a}, \op{a}^\dagger]=1$. We also define the associated position- and momentum-quadrature operators,
\begin{align}
	\op q \coloneqq \frac {1} {\sqrt 2} (\op a + \op a^\dag)
	\quad \text{and} \quad
	\op p \coloneqq \frac {-i} {\sqrt 2} (\op a - \op a^\dag),
\end{align}
respectively, so that $[\op q, \op p] = i$ with $\hbar = 1$, giving a numerical vacuum variance of $\langle \op q^2 \rangle = \langle \op p^2 \rangle = \frac 1 2$.

To discuss this circuit, we will need the standard definitions~\cite{bib:kok2010} for the squeezing operator
\begin{align}
\label{eq:BenBlows}
	\op{S}(\sqp) &\coloneqq \exp \left[\frac{1}{2}(\sqp^*\op{a}^2-\sqp \op{a}^{\dagger 2})\right],
\intertext{displacement operator}
\label{eq:ILikeToMoveItMoveIt}
	\qquad \op D(\alpha) &\coloneqq \exp (\alpha \op a^\dag - \alpha^* \op a),
\intertext{and displaced squeezed vacuum state}
	\ket{\alpha,\sqp} &\coloneqq \op D(\alpha) \op S(\sqp) \ket 0.
\end{align}
For brevity, when $\alpha=0$ we will simply write~$\ket \sqp \coloneqq \ket{0,\sqp}$ for a squeezed vacuum state. Note that when ${\sqp>0}$, $\ket \sqp$~is squeezed in~$\op q$ and anti-squeezed in~$\op p$, and vice versa for ${\sqp<0}$.

Using this notation, the joint state at (b) in Fig.~\ref{fig:CircuitModel} is
\begin{align}
\label{eq:stateatb}
\ket{\Psi_b }_{AO}= \sum_{\mathclap{m=-J}}^J d_{m,J}\ket{J,m}_A \otimes \ket{\sqp}_O.
\end{align}
The atomic ensemble and optical field are entangled with a controlled interaction %
that applies a displacement to the field proportional to the collective spin along the $z$-direction:
\begin{align} \label{eq:BigD}
	\op{D}_c(g) &\coloneqq \exp (-i g \op{p}\op{J}_z)
	= \op D(\gdisp \op J_z)
	= \op R_z(g\op p),
\end{align}
where we have defined
\begin{align}
	\gdisp &\coloneqq \frac {g} {\sqrt 2}
\end{align}
for brevity of notation and ease of converting between the Glauber~\cite{Glauber1963} and Weyl-Heisenberg displacement operators%
.

As shown in Eq.~\eqref{eq:BigD}, the interaction $\op D_c(g)$ can be thought of in two equivalent ways: (a)~${\op{D}_c(g) = \op D(\gdisp \op J_z)}$, which is
a position displacement of the optical mode controlled on $\op J_z$ of the ensemble, and (b)~${\op{D}_c(g) = \op R_z(g\op p)}$, which is
a $z$-rotation of the collective spin controlled on $\op p$ of the mode. The real parameter~$g$ represents the strength of the coupling in both of these interpretations. (We use the symbol~$g$ for this coupling strength because it will directly determine the spike spacing~$g$ of the resource state, which is discussed in Sec.~\ref{sec:GKPanatomy}.) 

$\op D_c(g)$ has the following effect on the components of the superposition in Eq.~\eqref{eq:stateatb}:
\begin{align} \label{eq:displacedsqueezed}
	\op D_c(g) \ket{J,m}_A\otimes \ket{\sqp}_O = \ket{J,m}_A \otimes \ket{\gdisp m,\sqp}_O. 
\end{align}
After the controlled displacement given by Eq.~(\ref{eq:BigD}), the joint state at (c) in Fig.~\ref{fig:CircuitModel} is
\begin{align} \label{eq:stateatc}
\ket{\Psi_c}_{AO} = \sum_{\mathclap{m=-J}}^J d_{m,J}\ket{J,m}_A \otimes \ket{ \gdisp m, \sqp}_O.
\end{align}

Finally, the collective spin is projectively measured along the $x$-direction as shown in (d) in Fig.~\ref{fig:CircuitModel}.  We express $\ket{\Psi_c}_{AO}$ in the basis of $x$-eigenstates: 
\begin{align}
\label{eq:psic}
	\ket{\Psi_c}_{AO} %
	&= \sum_{\mathclap{m,m'=-J}}^J  
	d_{m,J}d_{m,m'}\ket{J, m_x=m'}_A \otimes \ket{ \gdisp m, \sqp}_O,
\end{align}
where we used the fact that the matrix elements in Eq.~(\ref{eq:dmmd}) satisfy $\bra{J,m'}\op{R}_y (-\pi/2)\ket{J,m} = d_{m,m'}$. 
Then, measurement trivially collapses the $m'$ summation in Eq.~(\ref{eq:psic}) to a single term. That is, given measurement outcome $x$, the optical state is projected to
\begin{align}
\label{eq:projectedOpticalState}
 \ket{\psi^x_{d}}_O = \frac{1}{\sqrt{\mathcal{P}(x)}}\sum_{m=-J}^J d_{m,J}d_{m,x}\ket{\gdisp m, \sqp}_O.
\end{align}
The state is normalized by the probability of obtaining~$x$, 
\begin{align}
\label{eq:GKPmeasProb}
\mathcal{P}(x) = \sum_{\mathclap{m,m'}}d_{m,J}d_{m,x}d_{m',J}d_{m',x}e^{-\frac{1}{4}g^2e^{2\sqp}(m-m')^2},
\end{align}
as shown in Appendix~%
\ref{sec:GKPKraus}%
, noting that the squeezing parameter $\sqp$ is taken to be real. 

The conditional optical state can be written as a wave function in the position and momentum quadratures, %
respectively, as
\begin{subequations} \label{eq:psicondJ}
\begin{align}
\label{eq:psicondJpos}
    \psi_d^x(q) 
    = & \frac{e^{\sqp/2}}{\mathcal{P}(x)^{1/2}\pi^{1/4}} %
    \sum_{\mathclap{m=-J}}^J d_{m,J}d_{m,x} \exp\left[-\frac{(q-gm)^2}{2 e^{-2\sqp}}\right], \\
\label{eq:psicondJmom}
    \tilde \psi_d^x(p) 
    = & \frac{e^{-\sqp/2}}{\mathcal{P}(x)^{1/2}\pi^{1/4}} %
     \sum_{\mathclap{m=-J}}^J d_{m,J}d_{m,x}\exp\left[-igmp-\frac{p^2}{2e^{2\sqp}}\right].
\end{align}
\end{subequations}
For total spin~${J = 4}$ and~${J=\frac 9 2}$, the position- and momentum-space wave functions are plotted in Fig.~\ref{fig:resourceState}.%

Each measurement outcome~$x$ prepares a conditional position-space wave function, Eq.~(\ref{eq:psicondJpos}), which is a superposition of displaced squeezed states separated by~$g$
with amplitudes governed by the product of matrix elements~${d_{m,J}d_{m,x}}$. The momentum-space wave function, Eq.~(\ref{eq:psicondJmom}), is less straightforward, and we defer interpreting it until the next section wherein we restrict to ${x = \pm J}$.

\section{Resource state generation}
\label{sec:stategen}

\begin{figure*}[p]
\includegraphics[width=\columnwidth]{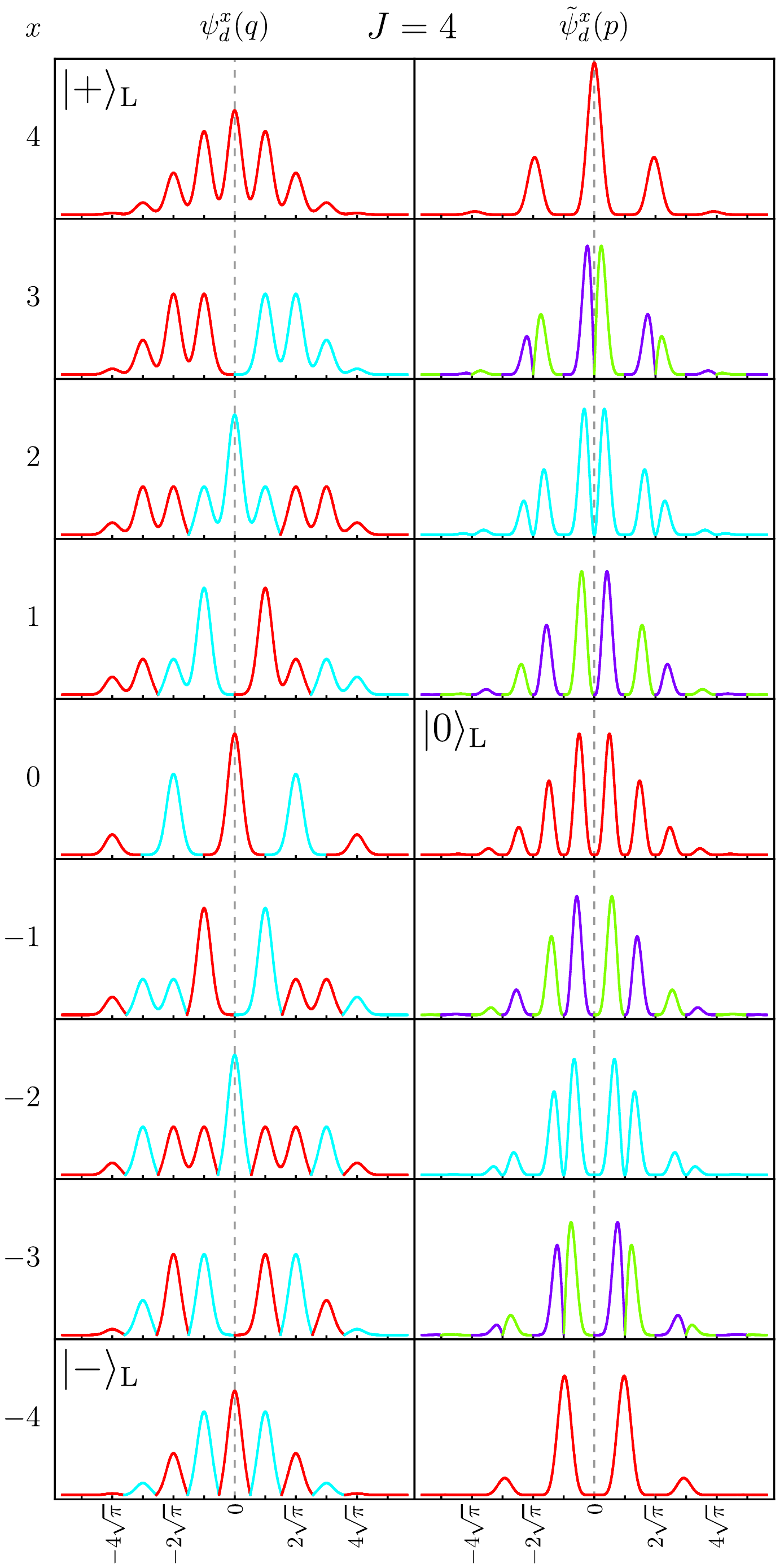}\hfill
\includegraphics[width=\columnwidth]{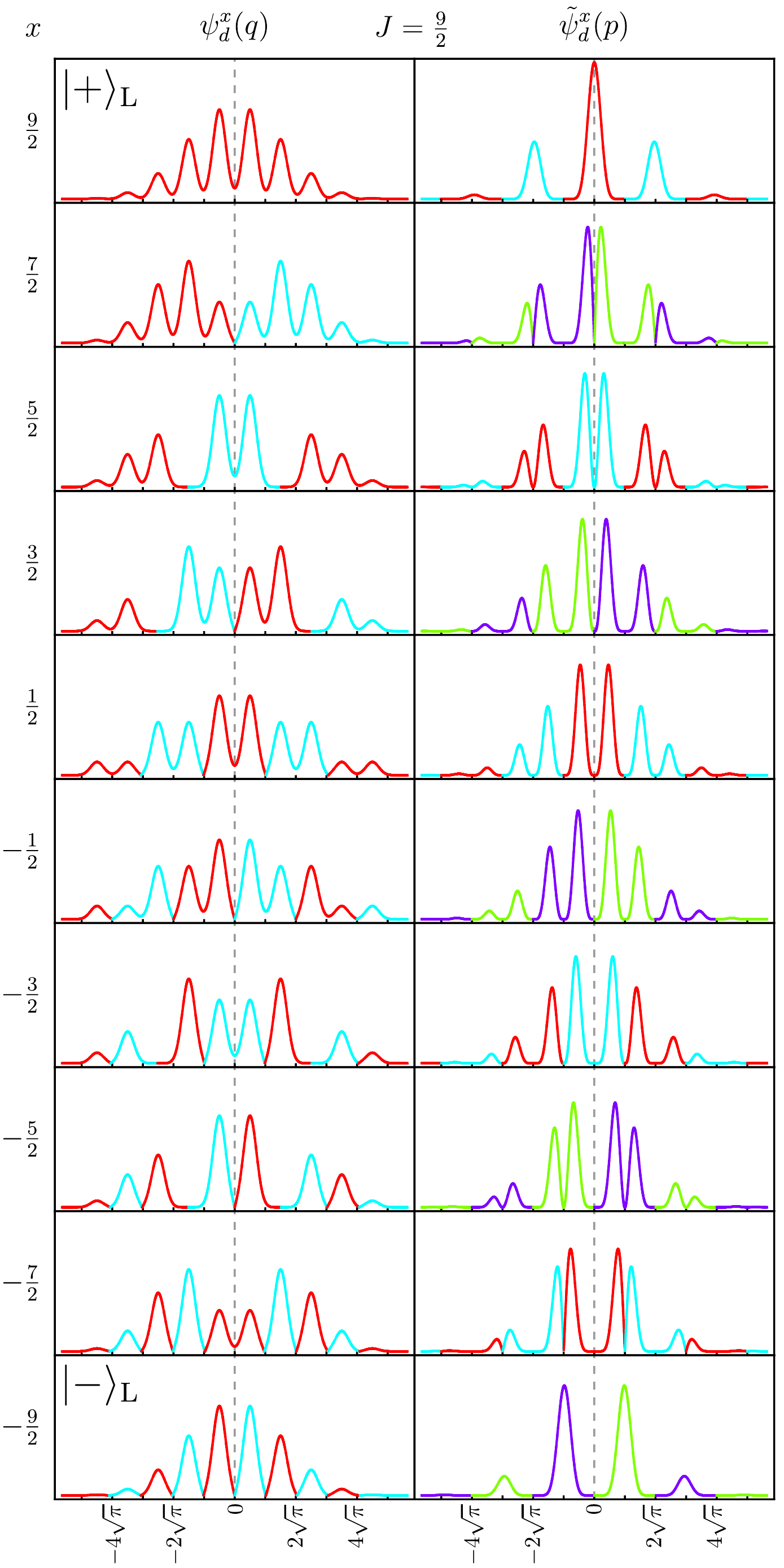}%
\caption{\label{fig:resourceState}Position- and momentum-space wave functions, Eqs.~(\ref{eq:psicondJ}), for the conditional optical state, $\ket{\psi_d^x}_O$, with total angular momentum $J=4$ and $J=\frac 9 2$, for each of the $2J+1$ measurement outcomes~$x$. Phase is indicated by color (grayscale) as in Fig.~\ref{fig:SymmResourceState}. When $x=\pm J$, the resulting state approximates the target state~$\ket \pm_\logic$ (after a position displacement, Eq.~\eqref{eq:shiftedlogical}, in the case of half-integer~$J$). To see this, compare $\psi_d^{\pm J}(q)$ with the position-space wave functions~$\psi_\pm(q)$ shown in the first row (for integer~$J$) and second row (for half-integer~$J$) of Fig.~\ref{fig:SymmResourceState}. We choose the initial squeezing of the optical mode and the coupling strength~$g$ to match the desired target state identified in Sec.~\ref{sec:requirements}. Specifically, $g=\sqrt\pi$ and $\sigma^2 = 2/\pi J$, corresponding to squeezing of 8.0~dB~($J=4$) and 8.5~dB~($J=\frac {9} {2}$), where $\text{(\#dB)} = 10 \log_{10}e^{2 \sqp}$. Also notice that the momentum-space wave function~$\tilde \psi_d^0(p)$, which only exists in the integer-$J$ case, represents the logical state~$\ket 0_\logic$ after a momentum displacement is applied. (Compare with $\psi_+(q)$ in the second row of Fig.~\ref{fig:SymmResourceState}, and see Sec.~\ref{sec:zerooutcome} for more details.) [Note for readers viewing in grayscale:~The relevant plots are the ones in the left-hand column of both tables, which are purely real, with red (darker) and cyan (lighter) indicating a relative phase of~$\pm 1$, respectively. Also relevant is the indicated~$\ket 0_\logic$ in the right-hand column of the $J=4$ table. It has a uniform phase of~$+1$. The other plots are presented merely for completeness and can be viewed in color online.]
}
\end{figure*}

The conditional optical states in Eq.~(\ref{eq:projectedOpticalState}) are useful as resource states (as defined in Sec.~\ref{sec:GKPtarget})
when three conditions are met:
\begin{enumerate}
\item The state's position-space wave function consists of a periodic comb of spikes with either uniform or alternating phase (corresponding to~$\ket\pm_\logic$, respectively).\vspace{1ex}
\item The spikes are separated in position by~$\sqrt \pi$. 
\item The embedded error of the state is symmetric---i.e., approximately equal in position and momentum.
\end{enumerate}
The first condition is met for spin-measurement outcomes $x=\pm J$. Tuning the coupling strength to $g=\sqrt{\pi}$ meets the second condition. The third condition is met by enforcing a constraint relating the 
optical squeezing~$\sqp$ and total spin~$J$. Below, we formalize these requirements and reveal the correspondence of the resource with the target GKP states~$\ket{\pm}_\logic$. The wave functions for the resource states appear in the top and bottom rows of both sides of Fig.~\ref{fig:resourceState}.

We begin with the first condition. Before enforcing any restrictions on $g$, the wave functions for the conditional optical states $\ket{\psi_d^{\pm J}}_O$ are%
\begin{subequations}
\label{eq:psiJ}
\begin{align}
\label{eq:psiqJ}
	\psi_d^{\pm J}(q) 
&
	=
	\frac{e^{\sqp/2}}{\mathcal{P}(\pm J)^{1/2}\pi^{1/4}}
	2^{-2J}  \sum_{m=-J}^J 
	\binom{2J}{J-m} (\pm 1)^{J+m}
\nonumber \\
&
\qquad\times
	\exp\left[-\frac{(q-gm)^2}{2 e^{-2\sqp}}\right],
	\\
\label{eq:psipJ}
	\tilde \psi_d^{\pm J}(p)
&
	=
	\frac{e^{-\sqp/2}}{\mathcal{P}(\pm J)^{1/2}\pi^{1/4}}
\nonumber \\
&
\qquad\times
	\exp\left(-\frac{p^2}{2e^{2\sqp}}\right)
	\choices
		{\cos^{2J}(g p/2)}
		{e^{iJ\pi}\sin^{2J}(g p/2)}
	,
\end{align}
\end{subequations}
where we have used the $d_{m,\pm J}$ matrix elements in Eq.~(\ref{eq:GKPmatrixElements}), and the top and bottom choices within the braces correspond to $\pm$, respectively.

The position-space wave function $\psi_d^{\pm J}(q)$ is a superposition of $2J+1$ Gaussians, each with variance
\begin{align} \label{eq:GKPvarQ}
\sigma_{q}^2=e^{-2\sqp},
\end{align}
separated by $g$, and distributed according to a binomial envelope arising from the binomial coefficients in Eq.~(\ref{eq:psiqJ}). The binomial coefficients may be approximated by a Gaussian using (see Appendix~\ref{app:analyticform})
\begin{align}
\label{eq:binomGauss}
	2^{-2J} \binom{2J}{J-m}
	&\approx \frac{1} {\sqrt{\pi J}} \exp \left(-\frac{m^2}{J}\right)
	,
\end{align}
which has variance~$J/2$. Since the spikes are located at $q=gm$ (instead of $q=m$), the binomial coefficients provide an approximately Gaussian envelope with variance
\begin{align} \label{eq:GKPvarEnvQ}
\sigma_{q,\text{env}}^2= \frac {g^2J} {2}.
\end{align}
Note, however, that the \emph{measured} variance of each spike is $\frac 1 2 \sigma_q^2$, and that of the envelope is $\frac 1 2 \sigma_{q,\text{env}}^2$, due to the need to square the wave function to get measured probabilities.

Similarly, the momentum-space wave function $\tilde \psi_d^{\pm J}(p)$ can be described as a product of a Gaussian envelope with variance
\begin{align} \label{eq:GKPvarEnvP}
\sigma_{p,\text{env}}^2=e^{2\sqp}
\end{align}
and a comb of approximately Gaussian peaks generated by $\cos^{2J}(g p/2)$ or $\sin^{2J}(g p/2)$ and hence separated by $2\pi/g$. The variance of the individual peaks is given by
\begin{align} \label{eq:GKPvarP}
    \sigma_{p}^2 &= \frac{2 \bigl[J^2\zeta(2,J)-1 \bigr]}{g^2J^2} %
	\approx
    \sigma_{q,\text{env}}^{-2} +O \left( \frac {1} {J^2} \right), 
\end{align}
where $\zeta(s,a)$ is the Hurwitz zeta function, as shown in Appendix~\ref{app:cosvariance}. As above, the measured variances of the spikes and envelope in momentum are $\frac 1 2 \sigma_p^2$ and $\frac 1 2 \sigma_{p,\text{env}}^2$, respectively. 

Under these approximations the wave functions are
\begin{subequations}
\label{eq:psiJapprox}
\begin{align}
\label{eq:psiqJapprox}
	\psi_d^{\pm J}(q) 
	&\simapprox
	e^{-q^2/2\sigma_{q,\text{env}}^2}
	\sum_{\mathclap{m=-J}}^J
	(\pm 1)^{J+m}
	\exp\left[-\frac{(q-gm)^2}{2 \sigma_q^2}\right],
	\\
\label{eq:psipJapprox}
	\tilde \psi_d^{\pm J}(p) 
	&\simapprox
	e^{-p^2/2\sigma_{p,\text{env}}^2}
	\sum_{\mathclap{s \in \choices {2\integers\phantom{+1}} {2\integers+1}}}
	e^{i s J \pi}
	\exp \left[-\frac{(p-s \pi / g)^2}{2 \sigma_p^2} \right]
	,
\end{align}
\end{subequations}
The position representation in Eq.~\eqref{eq:psiqJapprox} results from modifying the approximation in Eq.~\eqref{eq:binomGauss} by the replacement $m \mapsto q/g$ (since this is where the original function has nontrivial support). The momentum representation, Eq.~\eqref{eq:psipJapprox}, results from approximating $\cos^{2J} (gp/2)$ and $\sin^{2J} (gp/2)$ by an infinite series of appropriately sized Gaussian spikes centered on $p = s \pi/g$ (where $s$ is even for~$+$ and odd for~$-$), with each multiplied by the original function value at the center of the spike. 

For $g = \sqrt{\pi}$, the conditional optical states ($x=\pm J$) meet the second condition---i.e., they have equally spaced spikes in position and momentum and thus resemble %
the target states, Eqs.~\eqref{eq:targetq} and~\eqref{eq:targetp}. However, there is an important difference between the conditional optical states when they arise from spin measurements of integer~$J$ versus half-integer~$J$. 

Refer to the left side of Fig.~\ref{fig:resourceState}. For any integer~$J$ (case $J=4$ shown), the outcome ${x=+J}$ produces a resource state close to the target %
state ~$\ket{+}_\logic = \frac {1} {\sqrt{2}}(\ket{0}_\logic+\ket{1}_\logic)$ because it is an equally spaced comb with uniform phase on every spike. Similarly, the outcome ${x=-J}$ produces a resource state close to the target state ~$\ket{-}_\logic = \frac {1} {\sqrt{2}}(\ket{0}_\logic-\ket{1}_\logic)$ since the spikes have alternating sign within the superposition.

Now refer to the right side of that figure. For any half-integer~$J$ (case $J=\frac 9 2$ shown), the outcomes ${x=\pm J}$, respectively, produce a resource state close to a \emph{slightly shifted version} of the target %
states~$\ket{\pm}_\logic$. This is because the logical subspace, by definition~\cite{Gottesman2001}, has a spike centered at 0. This corresponds to the $m=0$ branch of the superposition in Eq.~\eqref{eq:projectedOpticalState}, which does not exist for half-integer~$J$. These states, therefore, must be displaced into the codespace by applying $\op D(\gdisp/2) = \exp(-i g\op p / 2)$ before they can be considered (approximate) logical GKP states.

Finally, the third condition is met when the embedded error is balanced in both quadratures ($\sigma_q^2 = \sigma_p^2$). 
Combining this requirement with the necessary coupling $g=\sqrt{\pi}$ and using Eqs.~\eqref{eq:GKPvarQ}, \eqref{eq:GKPvarEnvQ}, and~\eqref{eq:GKPvarP} gives the following relationship between squeezing~$\sqp$ and total spin~$J$:
\begin{align}
\label{eq:xisymmetric}
	J = \frac{2}{\pi} e^{2 \sqp} \qquad \Longleftrightarrow \qquad \sqp = \frac 1 2 \log \left( \frac {\pi J}{2} \right).
\end{align}
This relationship symmetrizes the spike and envelope variances:
\begin{align}
\label{eq:symmvars}
	\sigma_{(q,p)}^2 = e^{-2\sqp} = \frac {2} {\pi J} = \sigma^{-2}_{(q,p),\text{env}}.
\end{align}
Indeed, it is straightforward to verify that plugging these parameters into Eqs.~\eqref{eq:psiJapprox} and displacing the state if required (for half-integer $J$) produces wave functions of the form of Eqs.~\eqref{eq:targetq} and~\eqref{eq:targetp}, respectively, with $q_0 = \sqrt \pi/ 2$ for half-integer $J$. 

\begin{figure}[t]
\includegraphics[width=1.0\columnwidth]{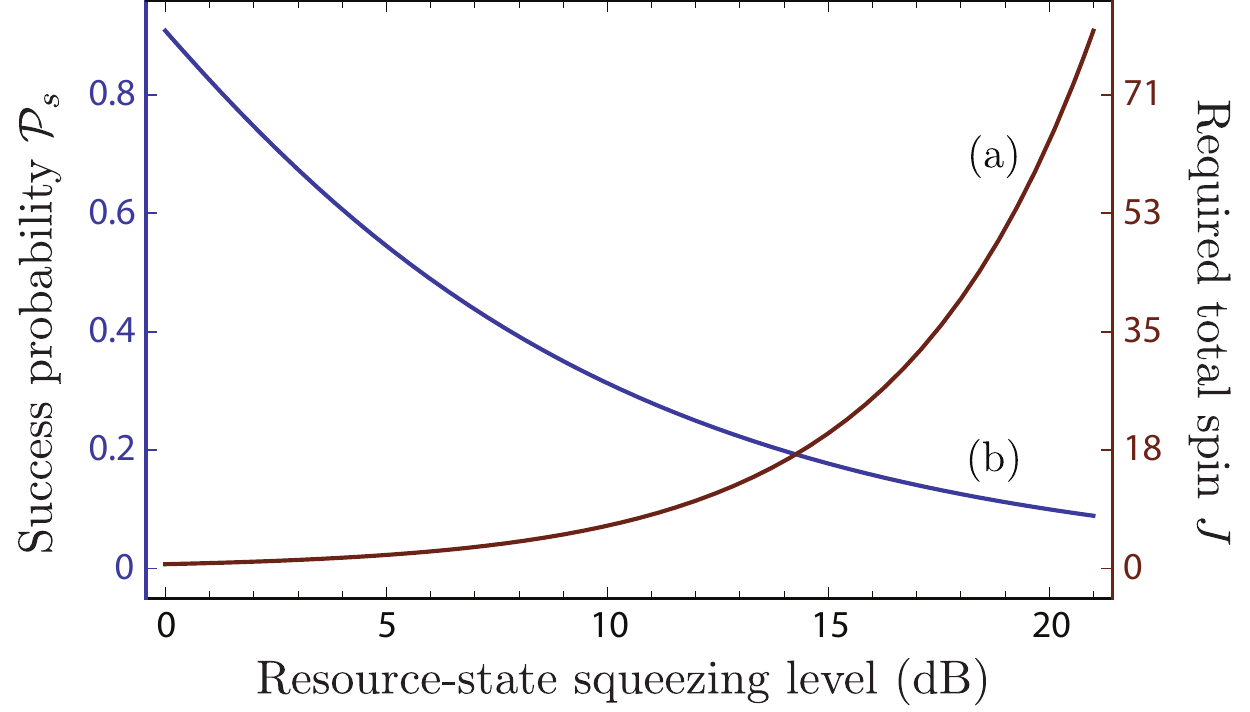}
\caption{(a)~Required total spin~$J$ [Eq.~\eqref{eq:Jfromsqueezing}] and (b)~success probability~$\mathcal P_s$ [Eq.~\eqref{eq:PsuccessfromdB}] for generating %
a resource state---i.e., an approximate GKP state satisfying the three conditions identified at the beginning of Sec.~\ref{sec:stategen}---with a given amount of squeezing. The resource state's squeezing level is defined in Eq.~\eqref{eq:squeezingofstate} and evaluated in Eqs.~\eqref{eq:ressqueezing}.
}
\label{fig:squeezing}
\end{figure}

Applying Eq.~\eqref{eq:squeezingofstate}, we can see that the squeezing of the state is equal to the optical squeezing (measured in~dB). Specifically,
\begin{subequations}
\label{eq:ressqueezing}
\begin{align}
&
	\text{(squeezing of the state in dB)}
\nonumber \\
&
\qquad
	=
	-10 \log_{10}e^{-2 \sqp} \approx 8.686\,\sqp
\\
&
\qquad
	=
	-10 \log_{10} \frac {2} {\pi J} \approx 1.9612 + 10 \log _{10} J
	.
\end{align}
\end{subequations}
We can invert the latter equation to obtain the required total spin~$J$ as a function of the resource state's squeezing (in~dB):
\begin{align}
\label{eq:Jfromsqueezing}
	J=\frac{2}{\pi} 10^{\text{(\#dB)}/10}.
\end{align}
The maroon curve~(a) in Figure~\ref{fig:squeezing} plots this relation for various levels of squeezing in the resource state. %
Total angular momentum ${J\approx 63.5}$, which corresponds to 20~dB of squeezing, is sufficient for fault-tolerant measurement-based quantum computing with continuous-variable cluster states~\cite{Menicucci:2014cx}, although more modest levels (e.g.,~${J\approx 20}$; 15~dB of squeezing) may still be useful in certain circumstances~\cite{Menicucci:2014cx}.

In summary, the resulting resource states produced by this method are 
\begin{align}
\label{eq:shiftedlogical}
	\ket{ \psiresource }
	\coloneqq
	\begin{cases}
		\op D\left( \gdisp /2 \right) \ket {\psi_d^{\pm J}} & \text{for half-integer~$J$,} \\
		\phantom{\op D\left( \gdisp /2 \right)} \ket {\psi_d^{\pm J}} & \text{for integer~$J$.}
	\end{cases}
\end{align} %
As these have a central spike at $q=0$ in their position-space wave function, equal spacing of spikes in position and momentum, and equal embedded error, they approximate well the target states $\ket{\pm}_\logic$, whose wave functions are shown in Fig.~\ref{fig:SymmResourceState}  (with $q_0 = \sqrt \pi/2$ for half-integer~$J$). \

Further, when the optical squeezing~$\sqp$ and total spin~$J$ are both increased while maintaining the functional relationship between them in Eq.~\eqref{eq:xisymmetric}, the resource states become better approximations to the target states, which themselves become better approximations to ideal GKP states. 

\subsection{Outcome \texorpdfstring{$x=0$}{x=0} for integer~\texorpdfstring{$J$}{J}}
\label{sec:zerooutcome}

When $J$~is an integer, there is another way to produce an approximate GKP state: the conditional state when~$x=0$. We choose not to focus too much on this option because (a)~the embedded error of the resulting state is different from that of the $x=\pm J$ outcomes, and (b)~it is not available for half-integer~$J$. Still, we detail this case here for completeness.

The fact that $x=0$ makes a useful state can be seen directly in Fig.~\ref{fig:resourceState} by noting that this outcome produces a momentum-space wave function~$\tilde \psi_d^0(p)$ that seems to resemble~$\psi_d^J(q)$, the position-space wave function for $x=J$. This would mean that it approximates the logical state~$\ket 0_\logic$ (since $p$ and~$q$ are related by a Fourier transform, which represents the logical Hadamard gate~\cite{Gottesman2001}). Let us formalize this resemblance.

Using the results from Appendix~\ref{app:analyticform}, we can immediately write the output wave functions for~$x=0$:
\begin{subequations}
\label{eq:x0psiJ}
\begin{align}
\label{eq:x0psiqJ}
	\psi_d^0(q)
&
	=
	\frac{e^{\sqp/2}}{\mathcal{P}(0)^{1/2}\pi^{1/4}}
	2^{-2J}
	{\binom{2 J}{J}}^{1/2}
\nonumber \\
&\qquad \times
	\sum_{\mathclap{m=-J}}^J
	\Re (i^{J+m})
	\binom{J}{\frac{J-m}{2}}
	\exp\left[-\frac{(q-gm)^2}{2 e^{-2\sqp}}\right]
	,
\\
\label{eq:x0psipJ}
	\tilde \psi_d^0(p) 
&
	=
	\frac{e^{-\sqp/2}}{\mathcal{P}(0)^{1/2}\pi^{1/4}}
	i^J 2^{-J}
	{\binom{2 J}{J}}^{1/2}
\nonumber \\
&\qquad \times
	\exp\left( -\frac{p^2}{2e^{2\sqp}} \right)
	 \sin^J (g p)
	.
\end{align}
\end{subequations}
Analogously to how we approximated Eqs.~\eqref{eq:psiJ} as Eqs.~\eqref{eq:psiJapprox}, we can approximate these as
\begin{subequations}
\label{eq:x0psiJapprox}
\begin{align}
\label{eq:x0psiqJapprox}
	\psi_d^0(q) 
	&\simapprox
	e^{-q^2/4\sigma_{q,\text{env}}^2}
	\sum_{\mathclap{m=-J}}^J
	\Re(i^{J+m})
	\exp\left[-\frac{(q-gm)^2}{2 \sigma_q^2}\right],
	\\
\label{eq:x0psipJapprox}
	\tilde \psi_d^0(p) 
	&\simapprox
	e^{-p^2/2\sigma_{p,\text{env}}^2}
\nonumber \\
&\quad \times
	\sum_{\mathclap{n \in \integers}}
	(-1)^{J n}
	\exp \left\{-\frac{\bigl[p-(n+\frac 1 2) \pi / g \bigr]^2}{\sigma_p^2} \right\}
	,
\end{align}
\end{subequations}
where the definitions of each of the $\sigma$'s is unmodified from above. It is evident that applying a momentum shift~$\op D(i \pi / 2 g \sqrt 2) = \exp(i \pi \op q / 2 g)$ will make $\psi_d^0 \simapprox \tilde \psi_+$ and $\tilde \psi_d^0 \simapprox \psi_+$, where the latter are defined in Eqs.~\eqref{eq:targetp} and~\eqref{eq:targetq}, respectively. As discussed above, this means that the $x=0$ output state (after the shift) approximates the logical state~$\ket 0_\logic$.

The most important difference between the wave functions in Eqs.~\eqref{eq:x0psiJapprox} and those in Eqs.~\eqref{eq:psiJapprox} is the embedded error. Specifically, the momentum-space spikes in Eq.~\eqref{eq:x0psipJapprox} are half the variance of the position-space spikes in Eq.~\eqref{eq:psiqJapprox}. Equivalently, the position-space envelope in Eq.~\eqref{eq:x0psiqJapprox} is twice the variance of the momentum-space envelope in Eq.~\eqref{eq:psipJapprox}.

Allowing for the outcome~$x=0$ in certain cases may prove beneficial for some applications, but for the reasons mentioned at the top of this subsection, we will not consider it further. Instead, we will focus on the results that are valid for all~$J$---i.e.,~the states resulting from the outcomes~$x=\pm J$.

\section{Success probability}
\label{sec:requirements}
Recall that the procedure for generating resource states described in Sec.~\ref{sec:lightensemble} is probabilistic, with the probability of spin-measurement outcome~$x$ given by~Eq.~\eqref{eq:GKPmeasProb}, repeated here for reference:
\begin{align}
\label{eq:GKPmeasProb2}
\mathcal{P}(x) = \sum_{\mathclap{m,m'}}d_{m,J}d_{m,x}d_{m',J}d_{m',x}e^{-\frac{1}{4}g^2e^{2\sqp}(m-m')^2}.
\end{align}
Valid outcomes for producing useful resource states are $x=\pm J$, which becomes [using Eq.~\eqref{eq:GKPmatrixElements}]
\begin{align}
	\mathcal{P}(\pm J)
&
	=
	2^{-4J}
	\sum_{\mathclap{m,m'}}
	(\pm 1)^{m+m'}
	\binom{2 J}{J-m} \binom{2 J}{J-m'}
\nonumber \\
&\qquad
	\times e^{-\frac{1}{4}g^2e^{2\sqp}(m-m')^2}
	.
\end{align}
Furthermore, for a symmetrically encoded state with balanced embedded error (see above), $g^2 e^{2\sqp} = \pi^2 J/2$, which means that, unless $J=0$, we really only need to consider the terms where $\abs{m-m'} \leq 1$  (the others being made negligible by the Gaussian). Therefore, for reasonably large $J$, we may approximate
\begin{align}
	e^{-\frac{1}{8}\pi^2 J(m-m')^2} &\approx \delta_{m,m'} + e^{-\frac{1}{8}\pi^2 J} (\delta_{m+1,m'} + \delta_{m,m'+1})
\end{align}
in the double sum (where $\delta$ is the Kronecker delta), which leads to
\begin{align}
	\mathcal{P}(\pm J)
	\approx
	\mathcal P_0(J) \pm 2 e^{-\frac{1}{8}\pi^2 J} \mathcal P_1(J),
\end{align}
where
\begin{align}
	\mathcal P_0(J) &\coloneqq 2^{-4J} \sum_{k=0}^{2J} \begin{pmatrix}2J \\ k\end{pmatrix}^2 = 2^{-4J} \begin{pmatrix}4J \\ 2J\end{pmatrix}, \\
	\mathcal P_1(J) &\coloneqq 2^{-4J} \sum_{k=0}^{2J-1} \begin{pmatrix}2J \\ k\end{pmatrix}  \begin{pmatrix}2J \\ k+1\end{pmatrix} = 2^{-4J} \begin{pmatrix}4J \\ 2J + 1\end{pmatrix}.
\end{align}
To obtain this expression, we 
shifted the (dummy) summation indices by~$J$ and used the Chu-Vandermonde identity to perform the sums.

Since either outcome $x = \pm J$ yields a useful state, the total success probability~$\mathcal P_s(J)$ is the sum of both cases. This means $\mathcal P_1(J)$ is irrelevant, giving (for~$J \ge \tfrac 1 2$)
\begin{align}
\label{eq:PsJapprox}
	\mathcal P_s(J)
	&\approx
	2 \times 2^{-4J} \begin{pmatrix}4J \\ 2J\end{pmatrix}
	.
\end{align}
Furthermore, we verified numerically that the following asymptotic approximation to Eq.~\eqref{eq:PsJapprox} (for large~$J$) is correct to within 1.3\% of the exact result [Eq.~\eqref{eq:GKPmeasProb2}] for all values of~$J \ge \tfrac 1 2$:
\begin{align}
\label{eq:PsuccessLargeJ}
	\mathcal P_s(J) &\approx
	\sqrt {\frac {2} {\pi J}} \left( 1 - \frac {1} {16 J} \right).
\end{align}
Using Eq.~\eqref{eq:Jfromsqueezing}, we can write this in terms of the squeezing of the state (measured in~dB):
\begin{align}
\label{eq:PsuccessfromdB}
	\mathcal P_s
&
	\approx
	10^{-\text{(\#dB)}/20}-\frac{\pi}{32} 10^{-3 \text{(\#dB)}/20}.
\end{align}
This is shown
as the blue curve~(b) in Fig.~\ref{fig:squeezing}. 

\begin{figure}[b]
\includegraphics[width=\columnwidth]{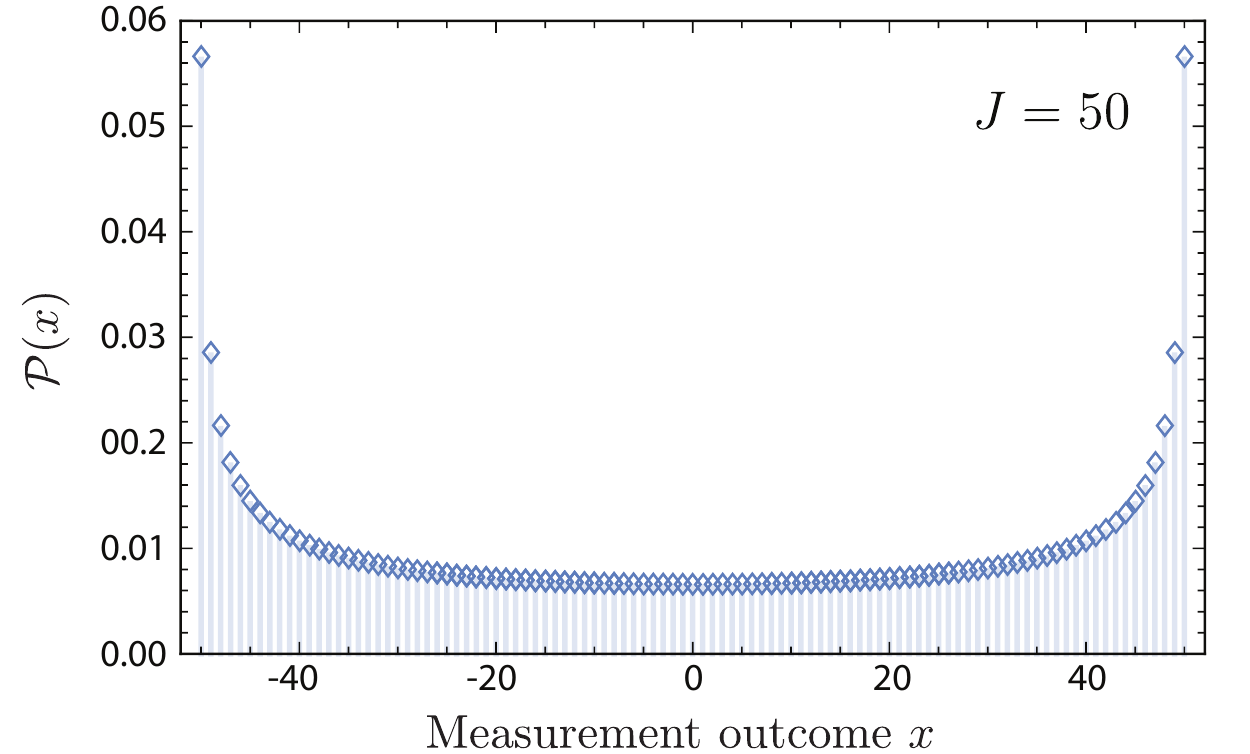}
\caption{The probability $\mathcal{P}(x)$ [Eq.~\eqref{eq:GKPmeasProb2}] of a particular spin-measurement outcome~$x$. The plot uses $J=50$ and a symmetric encoding with balanced embedded error (see the beginning of Sec.~\ref{sec:requirements}). Of the $2J+1$ possible outcomes, two of them---namely, $x=\pm J$, which also just happen to be the most probable---yield useful resource states. (In the case of integer~$J$, the $x=0$ case also produces a useful state, albeit with slightly different properties; see Sec.~\ref{sec:zerooutcome}.)
}
 \label{fig:hightails}
\end{figure}

Critical here is that the success probability scales as $\mathcal P_s \sim J^{-1/2}$ for large~$J$ (i.e., for high-quality resource states), whereas in a similar scheme proposed in Ref.~\cite{bib:Travaglione02} using a single spin-$\frac{1}{2}$ atom with repeated interactions and measurements, the success probability is approximately
\begin{align}
\label{eq:iteratedsucc}
	\mathcal P_{s,\text{iter}}(J) \sim 2^{-2J+1}.
\end{align}
This is exponential in $J = \NA/2$, where $\NA$ is the number of iterations of interaction with the atom. We obtained this formula from noting that the first measurement of this iterative method will succeed on either outcome%
, but on subsequent iterations, only one of the two outcomes will lead to a useful state~\cite{bib:Travaglione02}, and each outcome is approximately unbiased. (Contrary to what is claimed in that reference, these other outcomes are not even useful in the error-correction gadget~\cite{Gottesman2001} because they do not occupy the logical subspace.)

The advantage of our protocol can be explained by noting that the measurement is \emph{collective}, with outcomes arising solely within the spin-$J$ subspace
of ${\NA = 2J}$ \mbox{spin-$\tfrac 1 2$} particles%
. Thus, the total number of measurement outcomes is $2J+1$, and on top of that, the desired outcomes ($\pm J$) are the most likely ones---see Fig.~\ref{fig:hightails}.

In contrast, consider the procedure proposed in Ref.~\cite{bib:Travaglione02}, whereby the individual spin-$\frac{1}{2}$ atom is measured iteratively a total of $\NA=2J$ times. The measurement samples the full $2^{2J}$-dimensional Hilbert space, with the overwhelming majority of outcomes failing to produce the desired resource state. %
Because of the exponential scaling, such an iterative method has essentially no hope of producing a resource state with squeezing much higher than about 10~dB (for which $\mathcal P_{s,\text{iter}} \approx 3\times 10^{-4}$), whereas 10~dB of squeezing in our proposal ($J \approx 6.5$) has a 
much more reasonable
success probability $\mathcal P_s \approx 31\%$.

\section{Physical Implementation}
\label{sec:PhysicalImplementation}

The required coupling between an optical mode and a spin system, Eq.~\eqref{eq:BigD}, arises in the linearized limit of the dispersive Faraday (or Kerr-type) interaction \cite{bib:Hammerer2010}. The polarization of light is transformed due to circular birefringence in an ensemble of polarizable quantum scatterers \cite{bib:deutsch10}. For strong enough coupling, the interaction produces the symmetric encoding in Sec.~\ref{sec:requirements} and provides the foundation for a quantum non-demolition measurement of the spin required to produce approximate GKP states in the optical mode. 

\subsection{Faraday-based QND interaction} \label{sec:GKPinteraction}

We consider an optical field coupled to the collective spin formed by an ensemble of polarizable neutral atoms. Consider $\NA$ such atoms, each with effective spin $\op{j}_z = \frac{1}{2} \big(\ketbra{\uparrow}{\uparrow} - \ketbra{\downarrow}{\downarrow} \big)$ defined on metastable ground states $\{\uparrow, \downarrow \}$. The atoms couple to a common mode of light possessing two orthogonal linear polarizations, horizontal ($H$) and vertical ($V$), with respective annihilation operators $\op{a}_H$ and $\op{a}_V$. For an off-resonant field, the atoms and light become entangled via the dispersive Faraday interaction, $\op{U} = e^{-i\chi \op{S}_3 \op{J}_z}$, which describes a coupling of the collective atomic spin, $\op{J}_z = \sum_{n=1}^\NA \op{j}_z^{(n)}$, to the 3-component of the field's Stokes vector~\cite{bib:deutsch10}, 
	\begin{align} \label{Eq::Stokes}
		\op{S}_3 &= \frac{1}{2i} \big( \op{a}^\dagger_H \op{a}_V - \op{a}^\dagger_V \op{a}_H \big).
	\end{align}
The Faraday interaction generates a rotation of the Stokes vector around the 3-axis proportional to the atomic spin projection along $\op{J}_z$ with interaction strength characterized by the dimensionless, single-photon rotation angle~$\chi$.

The controlled displacement required for the resource-state generation, Eq.~(\ref{eq:BigD}), arises by preparing the $H$-mode of light in a coherent state with $\NL$ photons.
Making the linearization $\op{a}_H \rightarrow \sqrt{\NL}$, the Stokes operator in Eq.~\eqref{Eq::Stokes} becomes $\op{S}_3 \approx  \sqrt{ \NL/2 }\, \op{p},$ where $\op{p} = -i(\op{a}_V - \op{a}_V^\dagger )/\sqrt{2}$ is the momentum quadrature of the $V$-polarized mode~\cite{bib:Hammerer2010}. This linearized Faraday interaction generates the requisite controlled translations of $V$-mode photons, Eq.~(\ref{eq:BigD}), with effective coupling strength
	\begin{align} \label{Eq::Effectiveg}
		g = \chi \sqrt{ \frac {\NL} {2}  }.
	\end{align}

To begin the protocol, the state $\ket{\psi_b}_{AO}$ in Eq.~(\ref{eq:stateatb}) is prepared. Optical pumping intializes the atomic ensemble in a spin-coherent state along $x$, where $z$ is defined by the propagation direction of the light, and squeezed vacuum is separately prepared by pumping an optically nonlinear medium \cite{bib:Walls2008}.  Then, the Faraday interaction entangles the spins and light, producing the state $\ket{\psi_c}_{AO}$ in Eq.~(\ref{eq:stateatc}) when $g = \sqrt{\pi}$. This is achieved for interaction time $t = 2 \pi/\chi^2 \dot{N}_L$ for an $H$-polarized classical pump with photon flux $\dot{N}_L$.

\subsection{Projective spin measurement}
\label{sec:GKPmeasurement}

Once the optical field and atomic ensemble have become entangled, the collective spin state is projectively measured.
A single atomic spin may be measured by driving a cycling transition and detecting the resulting fluorescence~\cite{bib:Wineland11}. Concatenating with unitary transformations to cycle through the measurement basis, a projective measurement can be realized. With access to universal two-qubit control, this can be extended to many spins \cite{bib:Lloyd2001}.
However, since our resource-state protocol benefits from large collective spins composed of many constituent atoms, such a procedure is prohibitively time- and resource-consuming. %
We focus instead on a quantum nondemolition (QND) measurement of the collective spin. 
The collective spin is coupled via the same Faraday interaction to a second optical field that serves as a \emph{meter}. 
The meter experiences a spin-dependent polarization rotation that is measured via homodyne polarimetry~\cite{bib:deutsch10}. 
When the spin-meter coupling is strong, relative rotations from different projective $m$-values are distinguishable over the inherent shot noise in the meter, and the collective spin measurement becomes projective. 
This is indeed the same strong-coupling requirement for the GKP-state peaks to be sufficiently separated. 

To implement the measurement, the collective spin is first rotated into the $x$-basis with a $\pi/2$-pulse. The meter, propagating again along $z$, is initialized with $\NL$ photons in the $H$-mode. The quantum-mechanical $V$-mode is prepared in position-squeezed vacuum~$\ket{\sqp}_M$ (\emph{M} stands for `meter'), which has shot-noise variance $\sigma^2_M = e^{-2\sqp}$ for $q$-measurements. Polarimetry in the diagonal polarization basis implements a homodyne measurement of the position quadrature for $V$-mode photons, with the $H$-mode serving as the local oscillator~\cite{bib:baragiola14}.  
The degree to which the measurement is projective is determined by the distinguishability of the meter states, $\ket{\gdisp m, \sqp }_M = e^{ -i  g m \op{p} } \ket{\sqp}_M$, given by Eq.~(\ref{eq:Overlap}),
	\begin{align} \label{Eq::Resolution}
		\big| \overlap{ \gdisp m, \sqp}{\gdisp m',  \sqp} \big|^2 & = \exp \left[ - \left( \frac {g} {2 \sigma_M} \right)^2(m - m')^2 \right].
	\end{align}
In the limit that $(m - m')^2 \gg 4 \sigma_M^2/g^2 $, the meter states become orthogonal.  Thus, distinguishing neighboring eigenstates of $\op{J}_z$ requires
\begin{align}
	\frac {4 \sigma_M^2} {g^2} \ll 1,
\end{align}
with limits set by the characteristic coupling strength $g$ and the squeezed shot noise in the polarimeter, $\sigma_M^2$. Note that in contrast to the procedure for producing the GKP target states in Sec.~\ref{sec:requirements}, the spin-meter coupling here is not constrained by a specific value of $g$ for a given input squeezing $\sqp$ since the goal is only to produce distinguishable meter states.

Practical limitations on $\NL$ in both the preparation and the measurement stage arise for two related reasons.
First, the Faraday interaction is an effective description of the light-matter coupling when the quantum emitters remain far below saturation \cite{bib:Hammerer2010}. 
Second, increased $\NL$ precipitates more spontaneous photon scattering that spoils the QND interaction and measurement. 
Indeed, this has restricted QND spin squeezing to the Gaussian regime, far from a projective measurement. 
Overcoming the effects of decoherence necessitates that the coupling to the collective optical mode is large relative to all other modes.
This is characterized by the optical density per atom, $\eta \coloneqq \sigma_0/A_m$, the ratio of the resonant atomic scattering cross section $\sigma_0$ to the transverse mode area $A_m$. 
While typical optical densities per atom in free space, $\eta \sim 10^{-5}$~\cite{bib:baragiola14}, are far too weak for our purposes, those in engineered photonic environments such as photonic crystal waveguides can be much larger: $\eta \sim 1$ \cite{bib:goban14}. Operating near a band edge, ``slow light" can further enhance the interaction by several orders of magnitude~\cite{bib:goban14,bib:Hood2016}. 

A detailed study of optical pumping for atoms very near and strongly coupled to a waveguide is beyond the scope of our work; nevertheless, an estimate of the required coupling can be found from a free-space model for alkali atoms. 
Here, the Faraday rotation angle per photon per unit angular momentum given above is $\chi = \eta \Gamma / 2 \Delta $, for detuning~$\Delta$ and spontaneous emission rate~$\Gamma$~\cite{bib:deutsch10}.
To realize the coupling strength in Eq.~\eqref{Eq::Effectiveg} required for symmetric code states and an approximately projective measurement while limiting the number of free-space scattered photons, we find that for $\NL = 10^4$ and $\Delta = 500 \Gamma$ the required optical density per atom is $\eta \sim 25$, within the reach of near-term technology.
It may be possible to augment the atom-light coupling with an optical cavity~\cite{bib:mcconnell15} and to suppress the deleterious effects of photon scattering by judiciously selecting the effective spin within each atom~\cite{bib:Bohnet14}. 
Alternative fruitful avenues have opened in other physical architectures, where demonstrated strong coupling of ``artificial atoms" to photonic environments could provide the necessary interaction strength~\cite{bib:devoret13, bib:arnold15}. In such systems, Purcell enhancement of the total coupling rate has the potential to reduce the collective spin's susceptibility to other sources of noise.

\section{Conclusion}
\label{sec:conclusion}

The fact that photons do not interact with each other directly---only indirectly via material interactions---both helps and hinders optical approaches to quantum computing. On the one hand, this non-interaction means that room-temperature experiments are possible with low noise from the environment. On the other hand, some matter-based mechanism is required to get the photons to interact---even if this is just interaction with a detector followed by postselection~\cite{Knill2001}.

The Gottesman-Kitaev-Preskill~(GKP) encoding of qubits within light modes~\cite{Gottesman2001} offers easier Clifford operations and built-in robustness to small errors at the price of resource states that are challenging to produce.
This is analogous to how measurement-based quantum computing~\cite{Raussendorf2001,Menicucci2006} replaces the need for on-demand, controlled interactions with the generation of a resource state---in this case, a cluster state~\cite{Briegel2001,Zhang2006}---after which only (much easier) adaptive local measurements are required. In both cases, one trades away the requirement of many difficult interactions during the computation for the up-front single challenge of creating the required resource states, along with more modest requirements at run time.

Despite the challenges associated with their generation, 
there is every reason to take GKP states seriously as a viable encoding for quantum information. In addition to their potential for use with optical CV cluster states~\cite{Menicucci:2014cx}, which is what we have focussed on here, subsequent results have shown that circuit QED may also be a viable platform for implementing fault-tolerant CV measurement-based quantum computing. A recent proposal exists to produce GKP qubits in circuit QED~\cite{Terhal2016}, and scalable CV cluster states can be made in circuit QED, as well~\cite{Grimsmo2016}. Furthermore, GKP states have recently been shown to provide additional error-correction benefits when combined with a surface-code architecture~\cite{Terhal2015}, which could be used in both optics and circuit QED.

\begin{acknowledgments}
N.C.M.\ thanks Gerard Milburn for the initial suggestion to consider interactions with atomic ensembles. B.Q.B.\ thanks Gavin Brennen and Ivan Deutsch for helpful discussions. K.R.M.,\ B.Q.B.,\ and A.G.\ acknowledge the Australian Research Council Centre of Excellence for Engineered Quantum Systems (Project number CE110001013). N.C.M.\ acknowledges the Australian Research Council Centre of Excellence for Quantum Computation and Communication Technology (Project number CE170100012). N.C.M.\ was additionally supported by the Australian Research Council under grant number DE120102204 and by the U.S.\ Defense Advanced Research Projects Agency (DARPA) Quiness program under Grant No.\ W31P4Q-15-1-0004.
\end{acknowledgments}

\appendix

\section{Kraus-operator formalism}
\label{sec:GKPKraus}

The theory of generalized quantum measurements provides a convenient formalism (see, for example,~Ref.~\cite{bib:NielsenChuang00}) to describe the state-preparation protocol described above.  
The conditional optical state is expressed using a set of Kraus operators corresponding to the outcomes of a spin measurement. 

We begin with a product state between a spin of total angular momentum $J$ and an optical field, $\ket{\psi_b} = \ket{\phi}_A \otimes \ket{\psi}_O$, where the spin and field states are arbitrary. 
For measurement outcome $x$, the normalized, conditional field state is given by 
	\begin{align} \label{eq:CondState}
		\ket{\psi_d^x}_O = \frac{\op{A}_{x} \ket{\psi}_O}{ \sqrt{ \mathcal{P}(x) } },  
	\end{align}
where $\op{A}_{x}$ is a Kraus operator and $\mathcal{P}(x)$ is the probability of outcome $x$. %
Expressing the spin state in the $z$-basis, $\ket{\phi}_A = \sum_m c_m \ket{m}$, the Kraus operator associated with the controlled-displacement interaction in Eq.~(\ref{eq:BigD}) is
	\begin{align}
		\op{A}_{x}  &\coloneqq \bra{x} \op{D}_c(g) \ket{\phi}_A 
			 = \sum_{m=-J}^J c_m d_{m,x}  \,  e^{ -i g m \op{p} } . \label{eq:KrausOp} 
	\end{align}
This describes the conditional operation that implements a set of displacements on the field proportional to the initial spin distribution, $c_m$, and  outcome $x$. The probability of outcome $x$ is obtained by tracing over the initial field state,\allowdisplaybreaks[0]
	\begin{align} \label{eq:ProbKraus}
		\mathcal{P}(x) & = \mbox{Tr} \big[ \op{A}^{\dagger}_{x} \op{A}_{x}  \ketbra{\psi}{\psi} \big]  \\
		& = \sum_{m, m'= -J}^J c_m c_{m'}^* d_{m,x}  d_{m', x} \bra{\psi} e^{-i g(m-m')\op{p}}\ket{\psi}.\nonumber
	\end{align}	\allowdisplaybreaks

For a more general input state of light, $\op{\rho}_O$, which may be mixed due to losses and errors in the squeezing procedure, the conditional field state is given by
 	\begin{align} \label{Eq:CondRho}
		\op{\rho}^x_{d} = \frac{\op{A}_{x} \op{\rho}_O \op{A}^\dagger_{x}}{ \mathcal{P}(x)  } 
	\end{align}
with probability $\mathcal{P}(x) = \mbox{Tr}[ \op{A}_{x}^\dagger \op{A}_{x} \op{\rho}_O ]$.

\subsection*{Measurement probability}

In Sec.~\ref{sec:lightensemble} we consider the case of a position-squeezed input field, Eq.~(\ref{eq:BenBlows}), and an initial spin-coherent state corresponding to $c_m = d_{m,J}$ in Eq.~(\ref{eq:KrausOp}). The Kraus operator description, Eq.~(\ref{eq:CondState}), then gives the conditional field state, Eq.~(\ref{eq:projectedOpticalState}). 
The probability of outcome $x$ follows directly from Eq.~(\ref{eq:ProbKraus}),
\begin{align}
    \mathcal{P}(x) = 
     \sum_{m,m'} d_{m,J}d_{m,x}d_{m',J}d_{m',x}\braket{ \gdisp m  , \sqp}{ \gdisp m' , \sqp}.
\end{align}
To evaluate the expression, we note that for real~$\alpha$ and~$\sqp$, a displaced squeezed state can be written as 
\begin{eqnarray}
\ket{\alpha,\sqp} = \op{D}(\alpha)\op{S}(\sqp)\ket{0} = \op{S}(\sqp)\op{D}\big(\alpha e^{\sqp} \big)\ket{0}.
\end{eqnarray}
Then, the overlap between states of different displacements is calculated simply,
\begin{align}
\overlap{\alpha,\sqp}{\beta,\sqp} 
&= \exp\left[ -\smallfrac{1}{2}e^{2\sqp}(\alpha-\beta)^2\right], \label{eq:Overlap}
\end{align}
and the probability becomes %
\begin{align} \label{eq:Pxappendix}
\mathcal{P}(x) = \sum_{m,m'}d_{m,J}d_{m,x}d_{m',J}d_{m',x}e^{-\frac{1}{4}g^2e^{2\sqp}(m-m')^2}.
\end{align}

\begin{widetext}

\section{Analytic form of Wigner's (small) d-matrix elements}
\label{app:analyticform}

The matrix elements~$d^{(J)}_{m,m'}(\beta)$ of Wigner's (small) $d$-matrix can be summed analytically~\cite{Varshalovich1988} using Jacobi polynomials~$P_n^{(a,b)} (z)$:
\begin{align}
	d_{m,m'}^{(J)}(\beta) %
&
	= \sum_k (-1)^{k-m'+m}
	\frac{\sqrt{(J+m')!(J-m')!(J+m)!(J-m)!}}{(J+m'-k)!k!(J-k-m)!(k-m'+m)!}
	\left( \cos \frac {\beta} {2} \right)^{2j-2k+m'-m}
	\left( \sin \frac {\beta} {2} \right)^{2k-m'+m}
\\
&
	=
	\dsign_{m,m'}
	\left[
		\frac {s! (s+\mu+\nu)!} {(s+\mu)! (s+\nu)!}
	\right]^{\tfrac 1 2}
	\left( \sin \frac {\beta} {2} \right)^\mu
	\left( \cos \frac {\beta} {2} \right)^\nu
	P_s^{(\mu,\nu)} (\cos \beta)
	,
\end{align}
where
\begin{align}
	\mu &= \abs{m-m'}
&
	\nu &= \abs{m+m'}
&
	s &= J - \frac 1 2 (\mu + \nu),
&
	\dsign_{m,m'} =
	\begin{cases}
		1			& m' \ge m,
	\\
		(-1)^{m'-m}	& m' < m.
	\end{cases}
\end{align}
When we evaluate this for~$\beta = \pi/2$, this simplifies to
\begin{align}
	d_{m,m'}
	\coloneqq d_{m,m'}^{(J)}\left( \frac \pi 2 \right) %
&
	=
	\dsign_{m,m'}
	\left[
		\frac {(2s+\mu+\nu)!} {(s+\mu)! (s+\nu)!}
		\frac {s! (s+\mu+\nu)!} {(2s+\mu+\nu)!}
	\right]^{\tfrac 1 2}
	2^{s-J}
	P_s^{(\mu,\nu)} (0)
\\
&
	=
	\dsign_{m,m'}
	{\binom{2 J}{s+\mu }}^{\tfrac 1 2}{\binom{2 J}{s}}^{-\tfrac 1 2}
	2^{s-J}
	P_s^{(\mu,\nu)} (0)
	.
\end{align}
This is an equivalent analytic expression for the sum in Eq.~\eqref{eq:dmmd}.

The two special cases of this we need for analysis are $m'=\pm J$ and $m'=0$. The first ($m'=\pm J$) is rather trivial since $s=0$, for which the Jacobi polynomial~$P_0^{(a,b)}(z)=1$, leaving
\begin{align}
	d_{m,\pm J}
&
	=
	\frac {(\pm 1)^{J+m}} {2^J}
	{\binom{2 J}{J-m}}^{\tfrac 1 2}
	,
\end{align}
which agrees with Eq.~\eqref{eq:GKPmatrixElements}. The second case ($m'=0$) takes a bit more work to evaluate. In this case, $\mu = \nu = \abs m$ and $s = J - \abs m$, which gives
\begin{align}
\label{eq:dm0part1}
	d_{m,0}
&
	=
	\dsign_{m,0}
	{\binom{2 J}{J}}^{\tfrac 1 2}{\binom{2 J}{J-\abs{m}}}^{-\tfrac 1 2}
	2^{-\abs{m}}
	P_{J-\abs{m}}^{(\abs{m},\abs{m})} (0)
	.
\end{align}
The Jacobi polynomial in this case can be simplified by first expressing it in terms of an ordinary hypergeometric function using the relation~\cite{Bateman1953vol2}~(\S10.8)
\begin{align}
	P^{(a,b)}_n(z)
	=
	2^{-n} (z-1)^n
	\binom{b+n}{n} \,_2F_1 \left(-n,-a-n;b+1;\frac {z+1} {z-1} \right)
	.
\end{align}
When $b = a$ and $z=0$, this simplifies to
\begin{align}
	P^{(a,a)}_n(0)
&
	=
	(-2)^{-n}
	\binom{a+n}{n} \,_2F_1(-n,-a-n;a+1; -1)
\nonumber \\
&
	\overset{\text {(a)}}{=}
	(-2)^{-n}
	\binom{a+n}{n}
	\frac
		{\Gamma (1+a) \Gamma (1-\frac{n}{2})}
		{\Gamma (1-n) \Gamma (1+a+\frac{n}{2})}
\nonumber \\
&
	=
	(-2)^{-n}
	\binom{a+n}{\frac{n}{2}}
	\frac
		{\Gamma (1+\frac{n}{2}) \Gamma (1-\frac{n}{2})}
		{\Gamma (1+n) \Gamma (1-n)}
\nonumber \\
&
	\overset{\text {(b)}}{=}
	(-2)^{-n}
	\binom{a+n}{\frac{n}{2}}
	\cos \frac {n \pi} {2}
	.
\end{align}
In~(a) we have used Kummer's Theorem~\cite{Bailey1935}~(\S2.3) to reduce the hypergeometric function in this particular instance to an expression involving only gamma functions, and in~(b) we employed Euler's reflection lemma, which can be written $\Gamma(1+z) \Gamma(1-z) = \pi z \csc \pi z$. Plugging into Eq.~\eqref{eq:dm0part1}, we get
\begin{align}
	d_{m,0}
&
	=
	\dsign_{m,0}
	{\binom{2 J}{J}}^{\tfrac 1 2}{\binom{2 J}{J-\abs{m}}}^{-\tfrac 1 2}
	2^{-\abs{m}}
	(-2)^{-(J-\abs m)}
	\binom{J}{\frac{J-\abs m}{2}}
	\cos \frac {(J-\abs m)\pi} {2}
\nonumber \\
&
	=
	\Re (i^{J+m})
	2^{-J}
	{\binom{2 J}{J}}^{\tfrac 1 2}{\binom{2 J}{J-m}}^{-\tfrac 1 2}
	\binom{J}{\frac{J-m}{2}}
	.
\end{align}
Note that we have used complex notation in the final answer instead of a cosine (for brevity) and that no absolute values remain.

We can now calculate the products we need for the main text:
\begin{align}
\label{eq:dmJdmpmJ}
	d_{m,J}d_{m,\pm J}
&
	=
	\frac {(\pm 1)^{J+m}} {2^{2J}}
	{\binom{2 J}{J-m}}
	,
\\
	d_{m,J}d_{m,0}
&
	=
	\frac {\Re (i^{J+m})} {2^{2J}}
	{\binom{2 J}{J}}^{\tfrac 1 2}
	\binom{J}{\frac{J-m}{2}}
	.
\end{align}
Furthermore, using the Gaussian approximation~\cite{Spencer2014}~(\S5.4)
\begin{align}
	\binom{n}{k}
&
	\approx
	 2^{n} \sqrt{\frac{2} {\pi n}} \exp \left(-\frac{(n-2k)^2}{2n}\right)
	,
\end{align}
we can approximate these products as
\begin{align}
	d_{m,J}d_{m,\pm J}
&
	\approx
	\frac{(\pm 1)^{J+m}} {\sqrt{\pi J}} \exp \left(-\frac{m^2}{J}\right)
	,
\\
	d_{m,J}d_{m,0}
&
	\approx
	\Re (i^{J+m})
	\frac {\sqrt 2} {(\pi J)^{3/4}}
	\exp \left(-\frac {m^2}{2J} \right)
	,
\end{align}
valid for $m = o(J^{2/3})$---i.e., everywhere except at the far tails of the distribution (where it is very small anyway and thus irrelevant, assuming a reasonably sized~$J \gg \frac 1 2$). These are used in deriving the approximations in Eqs.~\eqref{eq:psiqJapprox} and~\eqref{eq:x0psiqJapprox}, respectively.
\vspace{1em}

\end{widetext}

\section{Variance of peaks in momentum representation}
\label{app:cosvariance}

The momentum representation of the resource states, Eq.~(\ref{eq:psipJ}), is a Gaussian envelope multiplying a comb generated by $\cos^{2J}(pg/2)$. We want to approximate each peak in the comb by a Gaussian by matching the peak's variance. Treating a single peak as a probability distribution we have
\begin{align}
    P(p) = \frac{g\Gamma(J+1)}{2\sqrt{\pi}\Gamma(J+ \frac 1 2)}\cos^{2J}(pg/2)
\end{align}
where the prefactors ensure the normalisation
\begin{align}
    \int_{-\pi/g}^{\pi/g}P(p)dp = 1.
\end{align}
The variance is then calculated in the usual way:
\begin{align}
    \sigma_p^2 = \langle p^2 \rangle - \langle p \rangle^2 =  \int_{-\pi/g}^{\pi/g}P(p)p^2dp.
\end{align}
Performing the integral we find
\begin{align}
    \sigma_p^2 = \frac{2(J^2\zeta(2,J)-1)}{g^2J^2},
\end{align}
where $\zeta(s,J)$ is the Hurwitz zeta function
\begin{align}
    \zeta(s,J) = \sum_{k=0}^\infty \frac{1}{(k+J)^{s}}.
\end{align}
For large $J$, $\zeta(2,J)\approx \frac{1}{J}+\frac{1}{2J^2}+O(\frac{1}{J^3})$ (see for example~\cite{bib:paris04}). Hence,
\begin{align}
    \sigma_p^2 \approx \frac{2}{g^2 J} + O\left( \frac {1} {J^2} \right) = \sigma_{q,\text{env}}^{-2} + O\left( \frac {1} {J^2} \right).
\end{align}

\bibliographystyle{bibstyleNCM_papers}
\bibliography{bibliography,MenicucciPapersRefs}

\begin{thebibliography}{10}

\bibitem{bib:NielsenChuang00}
M.~A. Nielsen and I.~L. Chuang, {\em Quantum Computation and Quantum
  Information} (Cambridge University Press, Cambridge, 2000).

\bibitem{Gottesman:2009ug}
D. Gottesman, ``{An Introduction to Quantum Error Correction and Fault-Tolerant
  Quantum Computation},'' arxiv:0904.2557v1 [quant-ph] (2009).

\bibitem{Gottesman2001}
D. Gottesman, A. Kitaev, and J. Preskill, ``{Encoding a Qubit in an
  Oscillator},'' Phys. Rev. A {\bf 64}, 012310 (2001).

\bibitem{Braunstein2005a}
S.~L. Braunstein and P. van Loock, ``{Quantum information with continuous
  variables},'' Rev. Mod. Phys. {\bf 77}, 513 (2005).

\bibitem{Vasconcelos:2010gb}
H.~M. Vasconcelos, L. Sanz, and S. Glancy, ``{All-optical generation of states
  for {\textquotedblleft}Encoding a qubit in an
  oscillator{\textquotedblright}},'' Opt. Lett., OL {\bf 35}, 3261 (2010).

\bibitem{bib:Travaglione02}
B.~C. Travaglione and G.~J. Milburn, ``Preparing encoded states in an
  oscillator,'' Phys. Rev. A {\bf 66}, 052322 (2002).

\bibitem{Pirandola:2004jo}
S. Pirandola, S. Mancini, D. Vitali, and P. Tombesi, ``{Constructing
  finite-dimensional codes with optical continuous variables},'' Europhys.
  Lett. {\bf 68}, 323 (2004).

\bibitem{Pirandola:2006gh}
S. Pirandola, S. Mancini, D. Vitali, and P. Tombesi, ``{Continuous variable
  encoding by ponderomotive interaction},'' Euro. Phys. J. D {\bf 37}, 283
  (2006).

\bibitem{Pirandola:2006bf}
S. Pirandola, S. Mancini, D. Vitali, and P. Tombesi, ``{Generating continuous
  variable quantum codewords in the near-field atomic lithography},'' J. Phys.
  B {\bf 39}, 997 (2006).

\bibitem{Terhal2016}
B.~M. Terhal and D. Weigand, ``Encoding a qubit into a cavity mode in circuit
  QED using phase estimation,'' Phys. Rev. A {\bf 93}, 012315 (2016).

\bibitem{Menicucci2006}
N.~C. Menicucci, P. van Loock, M. Gu, C. Weedbrook, T.~C. Ralph, and M.~A.
  Nielsen, ``{Universal Quantum Computation with Continuous-Variable Cluster
  States},'' Phys. Rev. Lett. {\bf 97}, 110501 (2006).

\bibitem{Yokoyama:2013jp}
S. Yokoyama {\it et~al.}, ``{Ultra-large-scale continuous-variable cluster
  states multiplexed in the time domain},'' Nature Photonics {\bf 7}, 982
  (2013).

\bibitem{Chen:2014jx}
M. Chen, N.~C. Menicucci, and O. Pfister, ``{Experimental Realization of
  Multipartite Entanglement of 60 Modes of a Quantum Optical Frequency Comb},''
  Phys. Rev. Lett. {\bf 112}, 120505 (2014).

\bibitem{Yoshikawa2016}
J.-i. Yoshikawa {\it et~al.}, ``{Generation of one-million-mode
  continuous-variable cluster state by unlimited time-domain multiplexing},''
  APL Photonics {\bf 1}, 060801 (2016).

\bibitem{Menicucci2011a}
N.~C. Menicucci, ``{Temporal-mode continuous-variable cluster states using
  linear optics},'' Phys. Rev. A {\bf 83}, 062314 (2011).

\bibitem{Wang:2014im}
P. Wang, M. Chen, N.~C. Menicucci, and O. Pfister, ``{Weaving quantum optical
  frequency combs into continuous-variable hypercubic cluster states},'' Phys.
  Rev. A {\bf 90}, 032325 (2014).

\bibitem{Alexander2016}
R.~N. Alexander, P. Wang, N. Sridhar, M. Chen, O. Pfister, and N.~C. Menicucci,
  ``One-way quantum computing with arbitrarily large time-frequency
  continuous-variable cluster states from a single optical parametric
  oscillator,'' Phys. Rev. A {\bf 94}, 032327 (2016).

\bibitem{Alexander:2014ew}
R.~N. Alexander, S.~C. Armstrong, R. Ukai, and N.~C. Menicucci, ``{Noise
  analysis of single-mode Gaussian operations using continuous-variable cluster
  states},'' Phys. Rev. A {\bf 90}, 062324 (2014).

\bibitem{Menicucci:2014cx}
N.~C. Menicucci, ``{Fault-Tolerant Measurement-Based Quantum Computing with
  Continuous-Variable Cluster States},'' Phys. Rev. Lett. {\bf 112}, 120504
  (2014).

\bibitem{Howe1988}
R. Howe, ``The oscillator semigroup,'' in {\em The Mathematical Heritage of
  Hermann Weyl}, \ Proc. Symp. Pure Math\ {\bf 48}, 61  (1988).

\bibitem{Varshalovich1988}
D. Varshalovich, A.~N. Moskalev, and V.~K. Khersonskii, {\em Quantum Theory Of
  Angular Momemtum} (World Scientific, Singapore, 1988).

\bibitem{bib:sakurai2011modern}
J.~J. Sakurai and J. Napolitano, {\em Modern quantum mechanics}
  (Addison-Wesley, 2011).

\bibitem{bib:kok2010}
P. Kok and B.~W. Lovett, {\em Introduction to Optical Quantum Information
  Processing} (Cambridge University Press, 2010).

\bibitem{Glauber1963}
R.~J. Glauber, ``Coherent and Incoherent States of the Radiation Field,'' Phys.
  Rev. {\bf 131}, 2766 (1963).

\bibitem{bib:Hammerer2010}
K. Hammerer, A.~S. S\o{}rensen, and E.~S. Polzik, ``Quantum interface between
  light and atomic ensembles,'' Rev. Mod. Phys. {\bf 82}, 1041 (2010).

\bibitem{bib:deutsch10}
I.~H. Deutsch and P.~S. Jessen, ``Quantum control and measurement of atomic
  spins in polarization spectroscopy,'' Optics Communications {\bf 283}, 681
  (2010).

\bibitem{bib:Walls2008}
D.~F. Walls and G.~J. Milburn, {\em {Quantum Optics}}, 2nd ed. (Springer,
  Berlin, 2008).

\bibitem{bib:Wineland11}
D. Wineland and D. Leibfried, ``Quantum information processing and metrology
  with trapped ions,'' Laser Phys. Lett. {\bf 8}, 175 (2011).

\bibitem{bib:Lloyd2001}
S. Lloyd and L. Viola, ``Engineering quantum dynamics,'' Phys. Rev. A {\bf 65},
  010101 (2001).

\bibitem{bib:baragiola14}
B.~Q. Baragiola, L.~M. Norris, E. Montano, P.~G. Mickelson, P.~S. Jessen, and
  I.~H. Deutsch, ``Three-dimensional light-matter interface for collective spin
  squeezing in atomic ensembles,'' Physical Review A {\bf 89}, 033850 (2014).

\bibitem{bib:goban14}
A. Goban {\it et~al.}, ``Atom--light interactions in photonic crystals,''
  Nature communications {\bf 5}, 3808 (2014).

\bibitem{bib:Hood2016}
J.~D. Hood {\it et~al.}, ``Atom--atom interactions around the band edge of a
  photonic crystal waveguide,'' Proceedings of the National Academy of Sciences
  {\bf 113}, 10507 (2016).

\bibitem{bib:mcconnell15}
R. McConnell, H. Zhang, J. Hu, S. Ćuk, and V. Vuletić, ``Entanglement with
  negative Wigner function of almost 3,000 atoms heralded by one photon,''
  Nature {\bf 519}, 439 (2015).

\bibitem{bib:Bohnet14}
J.~G. Bohnet, K.~C. Cox, M.~A. Norcia, J.~M. Weiner, Z. Chen, and J.~K.
  Thompson, ``Reduced spin measurement back-action for a phase sensitivity ten
  times beyond the standard quantum limit,'' Nature Photonics {\bf 8}, 731
  (2014).

\bibitem{bib:devoret13}
M.~H. Devoret and R.~J. Schoelkopf, ``Superconducting circuits for quantum
  information: an outlook,'' Science {\bf 339}, 1169 (2013).

\bibitem{bib:arnold15}
C. Arnold {\it et~al.}, ``Macroscopic rotation of photon polarization induced
  by a single spin,'' Nature communications 6 (2015).

\bibitem{Knill2001}
E. Knill, R. Laflamme, and G.~J. Milburn, ``{A scheme for efficient quantum
  computation with linear optics},'' Nature (London) {\bf 409}, 46 (2001).

\bibitem{Raussendorf2001}
R. Raussendorf and H.~J. Briegel, ``{A one-way quantum computer},'' Phys. Rev.
  Lett. {\bf 86}, 5188 (2001).

\bibitem{Briegel2001}
H.~J. Briegel and R. Raussendorf, ``{Persistent Entanglement in Arrays of
  Interacting Particles},'' Phys. Rev. Lett. {\bf 86}, 910 (2001).

\bibitem{Zhang2006}
J. Zhang and S.~L. Braunstein, ``{Continuous-variable Gaussian analog of
  cluster states},'' Phys. Rev. A {\bf 73}, 032318 (2006).

\bibitem{Grimsmo2016}
A.~L. Grimsmo and A. Blais, ``{Squeezing and quantum state engineering with
  Josephson traveling wave amplifiers},'' arXiv:1607.07908v1 [quant-ph] (2016).

\bibitem{Terhal2015}
B.~M. Terhal, ``Quantum error correction for quantum memories,'' Rev. Mod.
  Phys. {\bf 87}, 307 (2015).

\bibitem{Bateman1953vol2}
H. Bateman, {\em Higher Transcendental Functions} (McGraw-Hill, New York,
  1953), Vol.~II.

\bibitem{Bailey1935}
W.~N. Bailey, {\em Generalised Hypergeometric Series} (Cambridge, 1935).

\bibitem{Spencer2014}
J. Spencer, {\em {Asymptopia}}, Vol.~71 of {\em Student Mathematical Library}
  (American Mathematical Society, 2014).

\bibitem{bib:paris04}
R.~B. Paris, ``The Stokes phenomenon associated with the Hurwitz zeta function
  $\zeta(s, a)$,'' Proc. R. Soc. A {\bf 461}, 297 (2004).

\end{thebibliography}

\end{document}